\documentclass[preprint,showpacs,preprintnumbers,amsmath,amssymb]{revtex4}

\usepackage{graphicx}
\usepackage{dcolumn}
\usepackage{bm}
\usepackage{subfigure}
\usepackage{color}

\begin{document}
\bibliographystyle{prsty}

\preprint{}

\title{Bose-Bose mixtures in an optical lattice: First-order superfluid-insulator transition and elementary excitations}

\author{Takeshi Ozaki$^1$}
\email{j1209702@ed.kagu.tus.ac.jp}
\author{Ippei Danshita$^{2,3}$}
\author{Tetsuro Nikuni$^1$}
\affiliation{$^1$Department of Physics, Faculty of Science, Tokyo University of Science, Shinjuku, Tokyo 162-8601, Japan}
\affiliation{$^2$Yukawa Institute for Theoretical Physics, Kyoto University, Kyoto 606-8502, Japan}
\affiliation{$^3$Computational Condensed Matter Physics Laboratory, RIKEN, Wako, Saitama 351-0198, Japan}

\date{\today}

\begin{abstract}
We study ground-state phase diagrams and excitation spectra of Bose-Bose mixtures in an optical lattice by applying the Gutzwiller approximation to the two-component Bose-Hubbard model.
A case of equal hoppings and equal intra-component interactions for both components is considered.
Due to the existence of inter-component interaction, there appear several quantum phases, such as the superfluid, paired superfluid, and counterflow superfluid phases.
We find that the transition from superfluid (SF) to Mott insulator (MI) with even filling factors can be of the first order for a wide range of the chemical potential.
We calculate the excitation spectra as a useful probe to identify the quantum phases and the SF-to-MI transitions.
In the excitation spectra of the SF phase, there are two gapless modes and a few gapful modes, which respectively correspond to phase and amplitude fluctuations of the order parameters. At the SF-to-MI transition point, we show that the energy gaps of certain amplitude modes reach zero for the second-order transition while they remain finite for the first-order one.
Since the excitation spectrum can be measured by the Bragg scattering, we calculate the dynamical structure factor by using the linear response theory.
We consider three types of density fluctuations, and show that the density fluctuations are coupled to different excitation branches in different quantum phases.
\end{abstract}

\maketitle

\section{\label{sec1}Introduction}

Since the observation of the quantum phase transition from the superfluid (SF) to Mott insulator (MI), ultracold atoms trapped in an optical lattice have provided unique fields for studies of strongly correlated quantum matter~\cite{Adv_Phys_56_243, RMP_80_885}.
In particular, the experimental creation of mixtures of two types of bosons in optical lattices~\cite{Science_319_295, PRL_100_140401, PRA_77_011603, PRL_103_245301, PRL_105_045303} has opened up new possibilities to explore exotic quantum phases.
Previous studies have analyzed the two-species Bose-Hubbard model, which quantitatively describes systems of Bose-Bose mixtures in optical lattices, and predicted various quantum phases, including SF, MI, phase separation, paired superfluid (PSF), counterflow superfluid (CFSF), density wave, and supersolid phases~\cite{PRL_90_100401, PRL_90_150402, PRA_67_013606, PRA_68_053602, NJP_5_113, PRL_92_050402, PRL_92_030403, PLA_332_131, PRB_72_184507, PRL_98_190402, PRB_75_144510, PRA_75_053613, PRA_76_013604, PRA_77_015602, PRL_100_240402, PRA_79_011602, PRA_80_023619, PRB_80_245109, PRL_103_035304, PRA_81_063602, PRA_82_033630, PRB_82_180510, NJP_12_093008, PRA_84_041609, PRB_84_144411, JPSJ_81_024001}.
The PSF phase is a superfluid of composite particles that consist of two bosons of different types, and is present for attractive inter-component interaction~\cite{PRL_90_150402, PRL_92_050402, PRL_92_030403, PRB_75_144510, PRA_75_053613, PRA_79_011602, PRA_80_023619, PRL_103_035304, PRA_81_063602, PRB_82_180510, NJP_12_093008, PRA_84_041609}.
The CFSF is expected to emerge inside the Mott insulator for repulsive inter-component interaction, and is interpreted as a superfluid of composite particles consisting of a boson of one type and a hole of the other~\cite{NJP_5_113, PRL_90_100401, PRL_92_050402, PRB_75_144510, PRA_77_015602, PRA_80_023619, PRA_81_063602, NJP_12_093008, PRA_84_041609}.
In the PSF (CFSF) phase, only the in-phase (out-of-phase) motional degrees of freedom exhibits superfluidity, i.e. the motion is dissipationless, while the out-of-phase (in-phase) motion is prohibited.
Despite the intensive interest, these two exotic phases have not been experimentally observed so far.

Hu {\it et al.}~have reported the dipole oscillations in the SF, PSF, and CFSF phases in an one-dimensional optical lattice using the time-evolving block decimation method~\cite{PRA_84_041609}. 
In order to understand such a dynamical behavior, it is important to reveal the details of the excitations of Bose-Bose mixtures in an optical lattice.
The excitations of a one-component Bose-Hubbard model~\cite{PRB_40_546, PRL_81_3108} have been well understood from previous theoretical and experimental studies.
In the MI phase, the lowest two branches of the excitation spectrum are gapful and correspond to the particle- and hole-excitation modes~\cite{PRB_59_12184, PRA_63_053601, PRL_89_250404, PRA_73_033621}.
In the SF phase, the excitation spectrum has one gapless mode and gapful modes~\cite{PRL_89_250404, PRB_75_085106, PRL_100_050404, PRA_84_033602, PRB_84_174522, PRL_109_010401}.
The gapless mode corresponds to oscillations of the phase of the superfluid order parameter and is known as the Bogoliubov mode while the lowest gapful mode is regarded as the amplitude mode in the vicinity of the SF-to-MI transition at commensurate fillings.
It is also well-known that the amplitude mode becomes gapless at the critical point.
The excitation spectra have been experimentally detected via the Bragg spectroscopy~\cite{PRL_82_4569, PRL_88_120407, Nature_Phys_6_56, PRL_102_155301, PRL_106_205303, PRL_107_175302, PRL_109_055301} and the lattice-amplitude modulation~\cite{PRL_92_130403, PRL_93_240402, Nature_487_454} and been successfully used to characterize the two phases and the phase transitions.
Likewise, it is expected that elementary excitations can be used to distinguish the different phases and the SF-to-MI transitions in Bose-Bose mixtures.

In this paper, we study the quantum phase transitions and the excitation properties of Bose-Bose mixtures at zero temperature in an optical lattice using the Gutzwiller approximation (GA).
We assume equal hoppings and equal intra-component interactions for both components.
To describe the SF and MI phases, we directly apply the GA to the two-component Bose-Hubbard model.
However, this approach fails to capture the PSF and CFSF phases because these phases arise from the hopping of particle pairs or anti-pairs.
In order to analyze the phases, we use the effective Hamiltonian describing the degrees of freedom of pairs or anti-pairs within the second order-perturbation theory.
First, we determine the ground-state phase diagrams of this system, for which our main focus is placed on the first-order SF-to-MI transition.
Kuklov {\it et al.}~have performed Monte Carlo simulations on so-called two-component $J$-current model, which is a classical counterpart of the two-component Bose-Hubbard model at even total fillings, and found that the SF-to-MI phase transition at even total fillings can be of the first order for a certain region of the inter-component interaction~\cite{PRL_92_050402}.
More recently, Chen {\it et al.}~have applied a method based on the tensor product states to the two-component Bose-Hubbard model with attractive inter-component interaction and the hardcore constraint, and shown that the transition from incommensurate SF to MI can be of the first order as well~\cite{PRB_82_180510}.
In the present paper, we will show that this is also the case for the two-component softcore Bose-Hubbard model with repulsive inter-component interaction and that the region of the first-order transition is extended to a wide range of the chemical potential.
Second, in order to study the excitations, we extend the method for calculating excitation spectra of the Bose-Hubbard model \cite{EPL_72_162,PRA_84_033602} to the system of Bose-Bose mixtures.
We discuss the effect of inter-component interaction on the excitation spectra in the SF and MI phases.
We show that the gap of the in-phase amplitude mode vanishes at the SF-to-MI transition for odd total filling and another amplitude mode also becomes gapless at the second-order SF-to-MI transition for even total filling.
Finally, we investigate the response to density perturbation by applying the linear response theory to the lattice system.
In the SF, PSF, and CFSF phases, we calculate the dynamical structure factors regarding responses to the three types of density perturbation, namely one-component, in-phase, and out-of-phase fluctuations, in order to show that these phases can be identified by means of the Bragg scattering techniques.

This paper is organized as follows: in Sec.~\ref{sec2}, we explain formulations, which consist of the two-component Bose-Hubbard model, GA, effective Hamiltonian, linearized equation of motion and the linear response theory.
These formulations are used in Sec.~\ref{sec3}, \ref{sec4}, and \ref{sec5}.
In Sec.~\ref{sec3}, we obtain the phase diagrams for several parameters.
We discuss first-order phase transitions from SF to MI with even total fillings for repulsive inter-component interaction.
In Sec.~\ref{sec4}, we determine excitation spectra in the MI, SF, PSF, and CFSF phases.
We see several changes in the excitation spectra when hopping amplitude decreases from SF to MI.
In Sec.~\ref{sec5}, we apply the linear response theory to a Bose-Bose mixture system, and discuss the response to the Bragg scattering.
The conclusions are presented in Sec.~\ref{sec6}.

\section{\label{sec2}Formulation}
\subsection{\label{subsec:BHM}Two-component Bose-Hubbard model}
We consider a system of a $D$-dimensional hypercubic optical lattice loaded with a mixture of two different types of bosons, which can be two different hyperfine states~\cite{Science_319_295, PRL_103_245301, PRL_105_045303}, atomic species~\cite{PRA_77_011603}, and isotopes~\cite{PRA_84_011610}.
Assuming that the optical lattice is sufficiently deep compared to the chemical potential, we model the system by the two-component Bose-Hubbard model~\cite{PRL_81_3108},
\begin{align}
\label{eqn:Ham}
H=&\sum_{\alpha =1,2}\left[-t_{\alpha} \sum_{\langle i,j\rangle} (\hat{b}_{\alpha,i} ^\dag \hat{b}_{\alpha, j} + \hat{b}_{\alpha, j}^\dag \hat{b}_{\alpha, i}) + \frac{U_{\alpha}}{2}\sum_{i} \hat{n}_{\alpha, i}(\hat{n}_{\alpha, i}-1) - \mu_{\alpha} \sum_i\hat{n}_{\alpha, i}\right]  \nonumber \\
& +U_{12} \sum _i \hat{n}_{1,i}\hat{n}_{2,i},
\end{align}
where $\hat{b}_{\alpha, i} ^\dag$ $(\hat{b}_{\alpha, i})$ is the creation (annihilation) operator of the $\alpha (=1$ or $2)$ component at site $i$, $\hat{n}_{\alpha, i}\equiv \hat{b}_{\alpha, i} ^\dag \hat{b}_{\alpha, i} $ is the number operator, and $\langle i,j \rangle$ denotes the sum over the nearest neighbor sites.
$t_{\alpha}$, $U_{\alpha}$, and $\mu_{\alpha}$ are the hopping amplitude, the on-site intra-component interaction and the chemical potential, respectively; $U_{12}$ is the on-site inter-component interaction.
In experiments the magnitude of $U_{12}$ can be controlled by the Feshbach resonance~\cite{PRL_100_140401, PRL_100_210402, PRA_82_033609} and by the component-dependent optical lattice~\cite{PRL_105_045303, NJP_12_055013}.
We determine the quantum phases by the compressibility $\kappa = n_{\rm t}^{-1}\left(\frac{\partial n_{\rm t}}{\partial \mu_{\rm t}}\right)_{\mu_{\Delta}}$, the polarizability $\phi = \left(\frac{\partial  n_{\Delta}}{\partial \mu_{\Delta}}\right)_{\mu_{\rm t}}$, and the order parameters $\Phi_{\alpha, i} \equiv\langle \hat{b}_{\alpha, i} \rangle$, $\Phi_{i}^{\mathrm{p}}\equiv\langle \hat{b}_{1, i}\hat{b}_{2, i} \rangle$, and $\Phi_{i}^{\mathrm{c}}\equiv\langle \hat{b}_{1, i}\hat{b}_{2, i}^\dag \rangle$, where $n_{\rm t} = n_1 + n_2$, $n_{\Delta} = n_1 - n_2$, $\mu_{\rm t} = \mu_1 + \mu_2$, $\mu_{\Delta} = \mu_1 - \mu_2$, and $n_\alpha$ is the number of particles per site (filling factor) of the component $\alpha$.
The conditions for identifying each phase are summarized in Table~\ref{tab:phases}.
Notice that the incompressible phases, namely CFSF and one with no long-range order (LRO), are regarded as Mott insulators.
Instabilities toward the phase separation and the collapse are characterized by the conditions $\phi < 0$ and $\kappa < 0$, respectively.
In this paper, we consider the special case of $U_1=U_2=U>0$, $t_1=t_2=t$, $\mu_1=\mu_2=\mu$, and $|U_{12}|<U$.
If $U_{12}$ does not satisfy this condition, the mixtures lead to the phase separation for repulsive interaction or the collapse for attractive interaction~\cite{PRA_58_4836,JPSJ_81_024001}.

\begin{table}[tb]
\begin{center}
\begin{tabular}{| c | c | c | c | c | c |}
\hline
Phase & $\kappa$ & $\phi$ & $\Phi_{a}$ & $\Phi^{\rm p}$ & $\Phi^{\rm c}$ \\
\hline
\,\, SF \,\, & \,\, + \,\, & \,\, + \,\, & finite & finite & finite \\ 
\,\, PSF \,\, & \,\, + \,\, & \,\, 0 \,\, & 0 & finite & 0 \\
\,\, CFSF \,\, & \,\, 0 \,\, & \,\, + \,\, & 0 & 0 & finite \\
\,\, No LRO \,\, & \,\, 0 \,\, & \,\, 0 \,\,  & 0 & 0 & 0  \\
\hline
\end{tabular}
\end{center}
\caption{\label{tab:phases}
List of the quantities characterizing the different phases, namely the superfluid (SF), the paired superfluid (PSF), and the counterflow superfluid (CFSF), and the phase with no long-range order (LRO). $+$ or finite means that the corresponding quantity takes the positive or finite value. We drop the site index $i$ of the order parameters by assuming the spatial homogeneity.
}
\end{table}

\subsection{\label{subsec:GW}Gutzwiller approximation}
We use the GA to investigate the ground states and the excitation spectra of Eq.~(\ref{eqn:Ham}) at zero temperature.
We assume the Gutzwiller-type variational wave-function written as
\begin{align}
|\Psi\rangle=\prod_i\sum_{n_{1},n_{2}=0}^{\infty}f_{n_{1},n_{2}}^{(i)}(\tau)|n_{1},n_{2}\rangle_i,
\end{align}
where $|n_{1},n_{2}\rangle_i$ is the Fock state with $n_1$ and $n_2$ particles of the components $\alpha = 1$ and $2$ at site $i$, and the variational factor $f_{n_{1},n_{2}}^{(i)}$ has to satisfy the normalization condition $\sum_{n_{1},n_{2}}|f_{n_{1},n_{2}}^{(i)}|^2=1$.
One can obtain the equation of motion for $f_{n_{1},n_{2}}^{(i)}$ by imposing the stationary condition on the effective action $\int d\tau\langle \Psi |{\rm{i}}\hbar\frac{d}{d \tau}-\hat{H}|\Psi\rangle$, which leads to
\begin{align}
\label{eqn:GWe1}
{\rm{i}}\hbar\frac{df_{n_{1},n_{2}}^{(i)}}{d\tau}=&\left\{\sum_{\alpha}\left[ \frac{U}{2}n_{\alpha}(n_{\alpha}-1)-\mu n_{\alpha} \right]+U_{12}n_{1}n_{2}\right\}f_{n_{1},n_{2}}^{(i)}\nonumber \\
&-t\sum_{\langle j \rangle_i}\left( \Phi_{1, j}\sqrt{n_{1}}f_{n_{1}-1,n_{2}}^{(i)}+\Phi_{1, j}^{\ast}\sqrt{n_{1}+1}f_{n_{1}+1,n_{2}}^{(i)} \right)\\
&-t\sum_{\langle j \rangle_i}\left(\Phi_{2, j}\sqrt{n_{2}}f_{n_{1},n_{2}-1}^{(i)}+\Phi_{2, j}^{\ast}\sqrt{n_{2}+1}f_{n_{1},n_{2}+1}^{(i)}\right),\nonumber
\end{align}
where the superfluid order parameters for each component are $\Phi_{1, j}=\sum_{n_{1},n_{2}} f_{n_{1}-1,n_{2}}^{(j)\ast}\sqrt{n_{1}}f_{n_{1},n_{2}}^{(j)}$ and $\Phi_{2, j}=\sum_{n_{1},n_{2}} f_{n_{1},n_{2}-1}^{(j)\ast}\sqrt{n_{2}}f_{n_{1},n_{2}}^{(j)}$.
We define $\sum_{\langle j \rangle_i}$ as sum over neighboring sites of site $i$.
In the ground state, the wavefunction is stationary so that the coefficients can be written by
\begin{align}
\label{eqn:sta}
f_{n_{1},n_{2}}^{(i)}(\tau)=\tilde{f}_{n_{1},n_{2}}^{(i)}e^{-{\rm{i}}\tilde{\omega}_i\tau},
\end{align}
where $\tilde{f}_{n_{1},n_{2}}^{(i)}$ is the coefficient of stationary state that does not depend on time, and the phase factor $\tilde{\omega}_i$ is given by
\begin{align}
\hbar\tilde{\omega}_i=&\sum_{n_1,n_2}^{\infty}\left[ \frac{U}{2}n_1(n_1-1)-\mu n_1+\frac{U}{2}n_2(n_2-1)-\mu n_2 +U_{12}n_1n_2 \right]\left|\tilde{f}_{n_1,n_2}^{(i)}\right|^2\notag \\
&-t\sum_{\alpha}\sum_{\langle j \rangle_i}\left( \tilde{\Phi}_{\alpha, j}^\ast\tilde{\Phi}_{\alpha, i}+\tilde{\Phi}_{\alpha, j}\tilde{\Phi}_{\alpha, i}^\ast \right),
\end{align}
where $\tilde{\Phi}_{1, i}=\sum_{n_{1},n_{2}} \tilde{f}_{n_{1}-1,n_{2}}^{(j)\ast}\sqrt{n_{1}}\tilde{f}_{n_{1},n_{2}}^{(j)}$ and $\tilde{\Phi}_{2, i}=\sum_{n_{1},n_{2}} \tilde{f}_{n_{1},n_{2}-1}^{(j)\ast}\sqrt{n_{2}}\tilde{f}_{n_{1},n_{2}}^{(j)}$ are the superfluid order parameters for stationary state.
We calculate the ground state coefficients by solving the Gutzwiller equation (\ref{eqn:GWe1}) with the imaginary time propagation method \cite{PRA_76_023606,PRA_86_023623}.
More specifically, we follow the steps as shown below; (i) take the imaginary time $\tau^\prime={\rm{i}}\tau$ and set adequate initial coefficients, (ii) calculate the order parameter $\Phi_i$, (iii) put $\Phi_i$ into Eq.~(\ref{eqn:GWe1}), calculate new coefficients, (iv) iterate the steps (ii) and (iii) until the coefficients and the average energy converge.
In the actual calculation, the Hamiltonian matrix is truncated at finite values $n_{c1}$ and $n_{c2}$.
In the present work, we set the initial coefficient as a real number, and $n_{c1}=n_{c2}=n_{c}$ and use a sufficiently large $n_c$ so that the results do not depend on $n_c$.

\subsection{\label{subsec:eff}Effective Hamiltonian}
One cannot describe the PSF and CFSF phases with the equation of motion (3), because there the hopping term is treated as the first-order perturbation while the hopping of pairs (anti-pairs), which are the essential degrees of freedom in the PSF (CFSF) phase, is a process of the second order with respect to the hopping.
For instance, if one tries to use Eq. (3), these phases are not present in any regions of the phase diagrams.
Moreover, it fails to resolve dispersion of the gapless mode in the excitation spectra.
Hence, to describe the CFSF and PSF phases, we use an effective Hamiltonian that is restricted in the low-energy subspace of pairs or anti-pairs and account for tunneling of pairs or anti-pairs within the second-order perturbation theory.

In the limit of $t/U\ll 1$, the Hamiltonian of Eq.~(\ref{eqn:Ham}) can be written as $H= H_0+tV$ by treating the hopping term as a perturbation, where the non-perturbative Hamiltonian $H_0$ and perturbation $V$ are given as
\begin{align}
H_0&=\sum_{\alpha, i}\left[ \frac{U}{2}\hat{n}_{\alpha, i}(\hat{n}_{\alpha, i}-1)-\mu \hat{n}_{\alpha, i} \right]+U_{12}\hat{n}_{1, i}\hat{n}_{1, i},\\
V&=-\sum_{\alpha,\langle i,j \rangle}\left[\hat{b}_{\alpha, i}^\dag \hat{b}_{\alpha, j} + h.c.\right].
\end{align}
Using the second order perturbation theory, we can derive the effective Hamiltonian~\cite{PRL_90_100401,PRL_92_050402,PRA_75_053613,PRL_103_035304,NJP_12_093008}.
In this paper, for simplicity, we consider the case that the amplitude of the inter-component is close to intra-component interaction, $|U_{12}|\lesssim U$.

In the case of attractive inter-component interaction $U_{12}<0$, the low-energy subspace of pairs is described by a product over single-site Fock states with equal occupation of the two species.
For $0<(U+U_{12})/U\ll 1 $, the effective Hamiltonian is given by \cite{PRL_92_050402, PRA_75_053613, PRL_103_035304}
\begin{align}
\label{eqn:efH1}
H_\mathrm{eff}=H_0-\frac{2t^2}{U}\sum_{\langle i,j \rangle}\left[ \hat{n}_i(\hat{n}_j+1)+\hat{n}_j(\hat{n}_i+1)+ \hat{b}_{1i}^\dag\hat{b}_{2i}^\dag\hat{b}_{1j}\hat{b}_{2j}+\hat{b}_{1j}^\dag\hat{b}_{2j}^\dag\hat{b}_{1i}\hat{b}_{2i} \right].
\end{align}
The appropriate Gutzwiller wave-function for the effective Hamiltonian Eq.~(\ref{eqn:efH1}) is given by
\begin{align}
|\Psi_\mathrm{p}\rangle=\prod_i\sum_nf^{\mathrm{p}(i)}_{n}|n,n\rangle_i.
\end{align}
This wavefunction leads to the equations of motion,
\begin{align} 
\label{eqn:GWe2}
{\rm{i}}\hbar\frac{d}{d\tau}f^{\mathrm{p}(i)}_{n}=& \left[Un(n-1)-2\mu n+U_{12}n^2\right]f^{\mathrm{p}(i)}_n\nonumber\\
&-\frac{2t^2}{U}\sum_{\langle j \rangle_i}\left[ n(\bar{n}_j+1)+ \bar{n}_j(n+1)\right]f^{\mathrm{p}(i)}_n\nonumber\\
&-\frac{2t^2}{U}\sum_{\langle j \rangle_i}\left[ \Phi_j^\mathrm{p}n f^{\mathrm{p}(i)}_{n-1}+\Phi_j^{\mathrm{p}\ast} (n+1) f^{\mathrm{p}(i)}_{n+1} \right],
\end{align}
where $\Phi_i^\mathrm{p}= \sum_{n_i}\left(f^{\mathrm{p}(i)\ast}_{n_i-1}n_i f^{\mathrm{p}(i)}_{n_i}\right)$, and $\bar{n}_{j}\equiv\langle \hat{n}_{1,j}\rangle$ is the average particle number, which corresponds to the number of pairs in this phase.
In the ground state, we can describe the coefficients $f_{n_{1},n_{2}}^{\mathrm{p}(i)}(\tau)=\tilde{f}_n^{\mathrm{p}(i)}e^{-{\rm{i}}\tilde{\omega}_i^\mathrm{p}\tau}$, where $\tilde{\omega}_i^\mathrm{p}$ is given by
\begin{align}
\hbar\tilde{\omega}_i^\mathrm{p}\equiv&\sum_n\left\{ Un(n-1)-2\mu n +U_{12}n^2  -\frac{2t^2}{U}\sum_{\langle j \rangle_i}\left[ n(\bar{n}_j+1)+ \bar{n}_j(n+1)\right]\right\}\left| \tilde{f}_n^{\mathrm{p}(i)}\right|^2\notag\\
&-\frac{2t^2}{U}\sum_{\langle j \rangle_i}\left( \tilde{\Phi}_j^\mathrm{p}\tilde{\Phi}_i^{\mathrm{p}\ast}+\tilde{\Phi}_j^{\mathrm{p}\ast}\tilde{\Phi}_i^\mathrm{p} \right),
\end{align}
where $\tilde{\Phi}_i^\mathrm{p}= \sum_{n_i}\left(\tilde{f}^{\mathrm{p}(i)\ast}_{n_i-1}n_i \tilde{f}^{\mathrm{p}(i)}_{n_i}\right)$ is the pair superfluid order parameter for the stationary state.

On the other hand, for the repulsive inter-component interaction $U_{12}>0$, the low-energy subspace of particle-hole pairs is described over single-site Fock states with uniform total on-site occupation $n_{\rm t}$.
For $0<(U-U_{12})/U\ll 1$, the effective Hamiltonian is given by \cite{PRL_90_100401, NJP_12_093008}
\begin{align}
\label{eqn:efH2}
H_\mathrm{eff}= H_0-\frac{t^2}{U}\sum_{\alpha}\sum_{\langle i,j \rangle}\left[\hat{n}_{\alpha, i}(\hat{n}_{\alpha, j}+1)+\hat{n}_{\alpha, j}(\hat{n}_{\alpha, i}+1)\right] 
-\frac{2t^2}{U}\sum_{\langle i,j \rangle}\left( \hat{b}_{1i}^\dag \hat{b}_{2i}\hat{b}_{2j}^\dag \hat{b}_{1j}+\hat{b}_{1j}^\dag \hat{b}_{2j}\hat{b}_{2i}^\dag \hat{b}_{1i}\right),
\end{align}
and the appropriate Gutzwiller wave function is
\begin{align}
|\Psi_\mathrm{c}\rangle=\prod_i \sum_nf_n^{\mathrm{c}(i)}|n,n_{\rm t}-n\rangle_i.
\end{align}
From the calculations similar to the attractive case, we obtain the Gutzwiller equation,
\begin{align}
\label{eqn:GWe3}
{\rm{i}}\hbar\frac{d}{d \tau}f^{\mathrm{c}(i)}_n=& \left\{ \frac{U}{2}\left[2n^2+n_{\rm t}\left( n_{\rm t}-2n-1 \right)\right]-\mu n_{\rm t} +U_{12}n(n_{\rm t}-n)\right\}f_n^{\mathrm{c}(i)} \nonumber\\
&-\frac{2t^2}{U}\sum_{\langle j \rangle_i}\left[2n\bar{n}_j+n_{\rm t}(n_{\rm t}-n-\bar{n}_j+1)\right]f_n^{\mathrm{c}(i)}\nonumber\\
&-\frac{2t^2}{U}\sum_{\langle j \rangle_i}\left[ \Phi_{j}^\mathrm{c}\sqrt{n(n_{\rm t}-n+1)}f_{n-1}^{\mathrm{c}(i)}+\Phi_{j}^{\mathrm{c}\ast}\sqrt{(n+1)(n_{\rm t}-n)}f_{n+1}^{\mathrm{c}(i)}\right],
\end{align}
where $\Phi_{j}^\mathrm{c}=\sum_{n_j}\left[ f^{\mathrm{c}(j)\ast}_{n_j-1}\sqrt{n_j(n_{\rm{t}}-n_j+1)} f^{\mathrm{c}(j)}_{n_j}\right]$.
The ground state is described as $f_n^{\mathrm{c}(i)}=\tilde{f}_n^{\mathrm{c}(i)}e^{-\rm{i}\tilde{\omega}_i^{\rm{c}}\tau}$, where the phase factor $\tilde{\omega}_i^\mathrm{c}$ is given by
\begin{align}
\hbar\tilde{\omega}_i^\mathrm{c}\equiv&\sum_n\left[ \frac{U}{2}\left\{2n^2+n_{\rm t}\left( n_{\rm t} - 2n - 1 \right)\right\}-\mu n_{\rm t} +U_{12}n(n_{\rm t}-n) \right.\notag\\
&\left. -\frac{2t^2}{U}\sum_{\langle j \rangle_i}\left\{2n\bar{n}_j+n_{\rm t}(n_{\rm t}-n-\bar{n}_j+1)\right\}\right]\left| \tilde{f}_n^{\mathrm{c}(i)}\right|^2\notag\\
&-\frac{2t^2}{U}\sum_{\langle j \rangle_i}\left( \tilde{\Phi}_j^\mathrm{c}\tilde{\Phi}_i^{\mathrm{c}\ast}+\tilde{\Phi}_j^{\mathrm{c}\ast}\tilde{\Phi}_i^\mathrm{c} \right).
\end{align}
Here $\tilde{\Phi}_{j}^\mathrm{c}=\sum_{n_j}\left(\tilde{f}^{\mathrm{c}(j)\ast}_{n_j-1}\sqrt{n_j(n_{\rm{t}}-n_j+1)} \tilde{f}^{\mathrm{c}(j)}_{n_j}\right)$ is the counterflow superfluid order parameter for the stationary state.

\subsection{\label{sucsec:Bogo}Linearized equations of motion}
In this section, we derive the linearized equations of motion, which allow us to calculate the energies and the wave functions of elementary excitations. In Sec.~\ref{sec4}, we will use the derived equations to calculate the excitation spectra for several different phases.

We consider a small fluctuation from the stationary state given by
\begin{align}
f^{(i)}_{n_1,n_2}(\tau)=\left[\tilde{f}^{(i)}_{n_1,n_2}+\delta f^{(i)}_{n_1,n_2}(\tau)\right]e^{-{\rm{i}}\tilde{\omega}_i\tau}.
\end{align}
Assuming that the stationary state is homogeneous, we expand the fluctuation in terms of the plane wave,
\begin{align}
\label{eqn:delf}
\delta f_{n_1,n_2}^{(i)}(\tau)=\sum_\mathbf{k}\left(u_{n_1,n_2,\mathbf{k}}e^{{\rm{i}}\left(\mathbf{k}\cdot\mathbf{r}_i-\omega_\mathbf{k} \tau\right)}-v_{n_1,n_2,\mathbf{k}}^{\ast}e^{-{\rm{i}}\left(\mathbf{k}\cdot\mathbf{r}_i-\omega_\mathbf{k} \tau\right)}\right),
\end{align}
where $\mathbf{r}_i$ is the position vector of site $i$.
Substituting Eq.~(\ref{eqn:delf}) into Eq.~(\ref{eqn:GWe1}) and linearizing the equations with respect to the small fluctuations, we obtain
\begin{align}
\label{eqn:GW-bogo_1}
\begin{pmatrix}
A_{\mathbf{k}} & B_{\mathbf{k}}\\
-B_{\mathbf{k}}^\ast & -A_{\mathbf{k}}^\ast
\end{pmatrix}
\begin{pmatrix}\mathbf{u}_{\mathbf{k}}\\
\mathbf{v}_{\mathbf{k}}
\end{pmatrix}
=\hbar\omega_\mathbf{k}
\begin{pmatrix}
\mathbf{u}_{\mathbf{k}}\\
\mathbf{v}_{\mathbf{k}}
\end{pmatrix},
\end{align}
where $\mathbf{u}_{\mathbf{k}}$ and $\mathbf{v}_{\mathbf{k}}$ are $(n_c+1)^2$-dimensional vectors with the components $u_{n_1,n_2,\mathbf{k}}$ and $v_{n_1,n_2,\mathbf{k}}$, respectively.
The matrix elements of $A_{\mathbf{k}}$ and $B_{\mathbf{k}}$ are given as,
\begin{align}
\label{eqn:matA}
A_\mathbf{k}^{(n_1,n_2),(n_1^\prime,n_2^\prime)}\equiv&\left[\sum_{\alpha}\left(\frac{U}{2}n_{\alpha}(n_{\alpha}-1)-\mu n_{\alpha}\right)+U_{12}n_1 n_2-\hbar\tilde{\omega}_i\right]\delta_{n_1,n_1^\prime}\delta_{n_2,n_2^\prime}\nonumber\\
& -zt\left( \tilde{\Phi}_{1}\sqrt{n_1}\delta_{n_1-1,n_1^\prime}+\tilde{\Phi}_{1}^{\ast}\sqrt{n_1+1}\delta_{n_1+1,n_1^\prime} \right)\nonumber\\
& -zt\left( \tilde{\Phi}_{2}\sqrt{n_2}\delta_{n_2-1,n_2^\prime}+\tilde{\Phi}_{2}^{\ast}\sqrt{n_2+1}\delta_{n_2+1,n_2^\prime} \right) \\
\label{eqn:matB}
&-\epsilon(\mathbf{k})\left[C_{1,1}^1+C_{-1,-1}^1 + C_{1,1}^2+C_{-1,-1}^2 \right],\nonumber\\
B_\mathbf{k}^{(n_1,n_2),(n_1^\prime,n_2^\prime)}\equiv&\ \epsilon(\mathbf{k})\left[D_{1,-1}^1+D_{-1,1}^1 + D_{1,-1}^2+D_{-1,1}^2 \right],\\
C_{l,m}^1\equiv&\sqrt{n_1^\prime+(1+l)/2}\sqrt{n_1+(1+m)/2}\tilde{f}_{n_1^\prime+l,n_2^\prime}^{\ast}\tilde{f}_{n_1+m,n_2},\nonumber\\
C_{l,m}^2\equiv&\sqrt{n_2^\prime+(1+l)/2}\sqrt{n_2+(1+m)/2}\tilde{f}_{n_1^\prime,n_2^\prime+l}^{\ast}\tilde{f}_{n_1,n_2+m},\nonumber\\
D_{l,m}^1\equiv&\sqrt{n_1^\prime+(1+l)/2}\sqrt{n_1+(1+m)/2}\tilde{f}_{n_1^\prime+l,n_2^\prime}\tilde{f}_{n_1+m,n_2},\nonumber\\
D_{l,m}^2\equiv&\sqrt{n_2^\prime+(1+l)/2}\sqrt{n_2+(1+m)/2}\tilde{f}_{n_1^\prime,n_2^\prime+l}\tilde{f}_{n_1,n_2+m},
\end{align}
where we define $\epsilon(\mathbf{k})\equiv2t\sum_{l=1}^D \cos(k_l a)$, $a$ is the lattice spacing, and $D$ is the spatial dimension.
We can calculate the excitation spectrum of a given stationary state by diagonalizing Eq.~(\ref{eqn:GW-bogo_1}).

Since Eq.~(\ref{eqn:GW-bogo_1}) does not capture correctly the low-lying excitations of the PSF and CFSF phases, we also derive the linearized equations of motion from the effective Hamiltonians Eqs.~(\ref{eqn:efH1}) and (\ref{eqn:efH2}) using Eqs.~(\ref{eqn:GWe2}) and (\ref{eqn:GWe3}), respectively.
The linearized equations of motion for the effective Hamiltonians Eqs.~(\ref{eqn:efH1}) and (\ref{eqn:efH2}) take the same form as Eq.~(\ref{eqn:GW-bogo_1}).  For the former case, the basis vectors ${\bf u}_{\bf k}$ and ${\bf v}_{\bf k}$ are $(n_{c}+1)$-dimensional vectors with the components $u_{n,{\bf k}}$ and $v_{n,{\bf k}}$, and
the matrix elements are given by
\begin{align}
A_\mathbf{k}^{n,n^\prime}\equiv&\left\{ Un(n-1) -2\mu n +U_{12}n^2 -\frac{2zt^2}{U}\left[ n\left( \tilde{n} +1 \right)+\tilde{n}\left(n+1\right) \right] -\hbar\tilde{\omega}_i^\mathrm{p} \right\}\delta_{n,n^\prime}\notag\\
&-\frac{2zt^2}{U}\left[ \tilde{\Phi}^{\mathrm{p}\ast}\left( n+1 \right)\delta_{n+1,n^\prime}+\tilde{\Phi}^\mathrm{p}n\delta_{n-1,n^\prime} \right]\notag \\
&-\epsilon^\mathrm{p}(\mathbf{k})\left[ \left(n^\prime+1\right)\left(n+1\right)\tilde{f}_{n^\prime+1}^{\mathrm{p}\ast}\tilde{f}_{n+1}^\mathrm{p} + n^\prime n\tilde{f}_{n^\prime-1}^{\mathrm{p}\ast}\tilde{f}_{n-1}^\mathrm{p} \right],\\
B_\mathbf{k}^{n,n^\prime}\equiv&\ \epsilon^\mathrm{p}(\mathbf{k})\left[ n^\prime\left(n+1\right)\tilde{f}_{n^\prime-1}^\mathrm{p}\tilde{f}_{n+1}^\mathrm{p} + \left(n^\prime+1\right)n\tilde{f}_{n^\prime-1}^\mathrm{p}\tilde{f}_{n+1}^\mathrm{p} \right],
\end{align}
where we have defined $\epsilon^{\rm p}({\bf k})\equiv\frac{4t^2}{U}\sum_{l=1}^D\cos{(k_l a)}$.
For the latter case, the basis vectors ${\bf u}_{\bf k}$ and ${\bf v}_{\bf k}$ are $(n_{\rm t}+1)$-dimensional vectors with the components $u_{n,{\bf k}}$ and $v_{n,{\bf k}}$, and
the matrix elements are given by
\begin{align}
A_\mathbf{k}^{n,n^\prime}\equiv&\left\{ \frac{U}{2}\left[2n^2+n_{\rm t}\left( n_{\rm t} - 2n - 1 \right)\right]-\mu n_{\rm t} +U_{12}n(n_{\rm t}-n) \right.\notag\\
&\left.-\frac{2zt^2}{U}\left[ 2n\tilde{n}_j+n_{\rm t}(n_{\rm t}-n-\tilde{n}_j+2) \right] -\hbar\tilde{\omega}_i^\mathrm{c} \right\}\delta_{n,n^\prime}\notag\\
&-\frac{2zt^2}{U}\left[ \tilde{\Phi}^{\mathrm{c}\ast}\sqrt{\left(n+1\right)\left(n_{\rm t}-n\right)}\delta_{n+1,n^\prime}+\tilde{\Phi}^\mathrm{c}\sqrt{n\left(n_{\rm t}-n+1\right)}\delta_{n-1,n^\prime} \right]\notag \\
&-\epsilon^\mathrm{c}(\mathbf{k})\left[ \sqrt{\left(n^\prime+1\right)\left(n_{\rm t}-n^\prime\right)\left(n+1\right)\left(n_{\rm t}-n\right)}\tilde{f}_{n^\prime+1}^{\mathrm{c}\ast}\tilde{f}_{n+1}^\mathrm{c}\right.\notag\\
&\left. + \sqrt{n^\prime\left(n_{\rm t}-n^\prime+1\right)n\left(n_{\rm t}-n+1\right)}\tilde{f}_{n^\prime-1}^{\mathrm{c}\ast}\tilde{f}_{n-1}^\mathrm{c} \right],\\
B_\mathbf{k}^{n,n^\prime}\equiv&\ \epsilon^\mathrm{c}(\mathbf{k})\left[ \sqrt{n^\prime\left(n_{\rm t}-n^\prime+1\right)\left(n+1\right)\left(n_{\rm t}-n\right)}\tilde{f}_{n^\prime-1}^\mathrm{c}\tilde{f}_{n+1}^\mathrm{c}\right. \notag\\
&\left.+ \sqrt{\left(n^\prime+1\right)\left(n_{\rm t}-n^\prime\right)n\left(n_{\rm t}-n+1\right)}\tilde{f}_{n^\prime-1}^\mathrm{c}\tilde{f}_{n+1}^\mathrm{c} \right],
\end{align}
where we define $\epsilon^{\rm c}({\bf k})\equiv\frac{4t^2}{U}\sum_{l=1}^D\cos{(k_l a)}$.

\subsection{\label{subsec:LRT}Linear response theory}
In this section, we briefly present the linear response theory applied to the equations of motion, namely Eqs.~(\ref{eqn:GWe1}), (\ref{eqn:GWe2}), and (\ref{eqn:GWe3}), which describes Bose-Bose mixtures in an optical lattice. 
More detailed derivations are shown in Appendix A. 
In Sec.~\ref{sec5}, we will use the linear response theory to calculate the dynamical structure factors for three types of density fluctuations.

We consider the time-dependent perturbation of the following form: 
\begin{align}
\label{eqn:a1}
\hat{H}_\mathrm{pert}(\tau)=-\sum_i\left( \lambda_i\hat{G}_ie^{-{\rm{i}}\omega \tau}e^{\eta \tau}+\lambda_i^\ast\hat{G}_i^\dag e^{{\rm{i}}\omega \tau}e^{\eta \tau} \right),
\end{align}
where $\lambda_i$ is the strength of external field that will be taken to be sufficiently small, $\eta$ is a small constant describing the adiabatic switching on of the perturbation at $t\to-\infty$, and $\hat{G}_i$ is the local operator given by
\begin{align}
\label{eqn:a2}
\hat{G}_i=\sum_{n_1,n_2,m_1,m_2}G_{m_1,m_2,n_1,n_2}|m_1,m_2\rangle_i\langle n_1,n_2 |_i.
\end{align}
The external field can be expanded in terms of plane waves as $\lambda_i=\sum_\mathbf{k}\lambda_{\mathbf{k},\omega}e^{{\rm{i}}\mathbf{k}\cdot\mathbf{r}_i}$.
The fluctuation of the expectation value of an local operator of physical interest $\hat{F}_i\equiv\sum_{n_1,n_2,m_1,m_2}F_{m_1,m_2,n_1,n_2}|m_1,m_2\rangle_i\langle n_1,n_2 |_i$ is given by
\begin{align}
\delta\langle\hat{F}_i\rangle=\sum_{\mathbf{k}}\left[ \chi_{\hat{F},\hat{G}}(\mathbf{k},\omega)e^{-{\rm{i}}\omega \tau}e^{\eta \tau}+ \chi_{\hat{F},\hat{G}^\dag}(\mathbf{k},-\omega)e^{{\rm{i}}\omega \tau}e^{\eta \tau} \right]e^{{\rm{i}}\mathbf{k}\cdot\mathbf{r}_i}\lambda_{\mathbf{k},\omega}.
\end{align}
Within the GA, the response function $\chi_{\hat{F},\hat{G}}(\mathbf{k},\omega)$ is given by (for a detailed derivation, see Appendix~\ref{sec:appA})
\begin{align}
\label{eqn:RF}
\chi_{\hat{F},\hat{G}}(\mathbf{k},\omega)=-\frac{1}{\hbar}\sum_\nu\left[ \frac{\langle 0|\hat{F}|\nu\rangle\langle\nu|\hat{G}|0\rangle}{\omega+{\rm{i}}\eta-\omega_{\mathbf{k},\nu}} - \frac{\langle 0|\hat{G}|\nu\rangle\langle\nu|\hat{F}|0\rangle}{\omega+{\rm{i}}\eta+\omega_{\mathbf{k},\nu}} \right],
\end{align}
where $|0\rangle$ is the ground state, and $|\nu\rangle$ is the $\nu$~th excited state.
The matrix element of the response function is defined as
\begin{align}
\label{eqn:LR_chi}
\langle0|\hat{O}|\nu\rangle\equiv\sum_{n_1,n_2}\left(\tilde{f}_{n_1,n_2}^\ast O_{n_1,n_2,n_1,n_2} u_{n_1,n_2,\mathbf{k}}^{(\nu)}-v_{n_1,n_2,\mathbf{k}}^{(\nu)} O_{n_1,n_2,n_1,n_2} \tilde{f}_{n_1,n_2} \right),\notag\\
\langle\nu|\hat{O}|0\rangle\equiv\sum_{n_1,n_2}\left(u_{n_1,n_2,\mathbf{k}}^{(\nu)\ast} O_{n_1,n_2,n_1,n_2} \tilde{f}_{n_1,n_2}-\tilde{f}_{n_1,n_2}^\ast O_{n_1,n_2,n_1,n_2} v_{n_1,n_2,\mathbf{k}}^{(\nu)\ast} \right),
\end{align}
where $u_{n_1,n_2,\mathbf{k}}^{(\nu)}$ and $v_{n_1,n_2,\mathbf{k}}^{(\nu)}$ are the solutions of the linearized equation of motion (\ref{eqn:GW-bogo_1}).
In the calculations of Sec.~\ref{sec5}, we set $\eta/U=10^{-2}$ for the SF phase and $\eta/U=10^{-4}$ in the PSF and CFSF phases.
We choose the operators $\hat{F}=\hat{G}$ and calculate the response function using the results of Sec.~\ref{subsec:Exc_SF}.
We obtain the dynamical structure factor $S_{\hat{F}}(\mathbf{k},\omega)$ from following relation \cite{pita_stri},
\begin{align}
\label{eqn:LR_SQO}
S_{\hat{F}}(\mathbf{k},\omega)={\rm{Im}}\left(\chi_{\hat{F},\hat{F}}(\mathbf{k},\omega)/\pi\right).
\end{align}
Note that this formulation is valid only at zero temperature and $S_{\hat{F}}(\mathbf{k},\omega)$ vanishes for $\omega<0$.
We normalize the dynamical structure factor as $\bar{S}_F(\mathbf{k},\omega)=S_F(\mathbf{k},\omega)/S_F(\mathbf{k})$ by using the static structure factor given by
\begin{align}
S_{\hat{F}}({\bf k})\equiv\int d\omega S_{\hat{F}}({\bf k},\omega).
\end{align}
%

\section{\label{sec3}Ground states and first order transitions}
In this section, we use the GA described in the previous section to obtain the phase diagrams in the $(zt/U, \mu/U)$ plane for equal hoppings, equal intra-component interactions, and several values of $U_{12}/U$, where $z$ is the coordination number.
Since we will investigate properties of elementary excitations in different quantum phases in Secs.~\ref{sec4} and \ref{sec5}, it is useful to locate these phases beforehand.
In previous studies, similar phase diagrams have been calculated by using mean-field theories~\cite{PRA_67_013606, PLA_332_131} and the strong-coupling expansion techniques~\cite{PRA_82_033630}.
However, the Mott insulators with odd total fillings have not been addressed in these studies. Moreover, while the first-order SF-to-MI transition was predicted to occur at even total fillings~\cite{PRL_92_050402}, it has not been confirmed that the first-order property remains for the transition between MI and incommensurate SF except in the case of hardcore bosons with attractive inter-component interaction~\cite{PRB_82_180510}.
Below we will address these points.
   
\subsection{\label{subsec:PHA}Phase diagrams}
\begin{figure}[tb]%
\begin{center}
\begin{tabular}{cc}
\subfigure[]{\includegraphics[width=7.2cm]{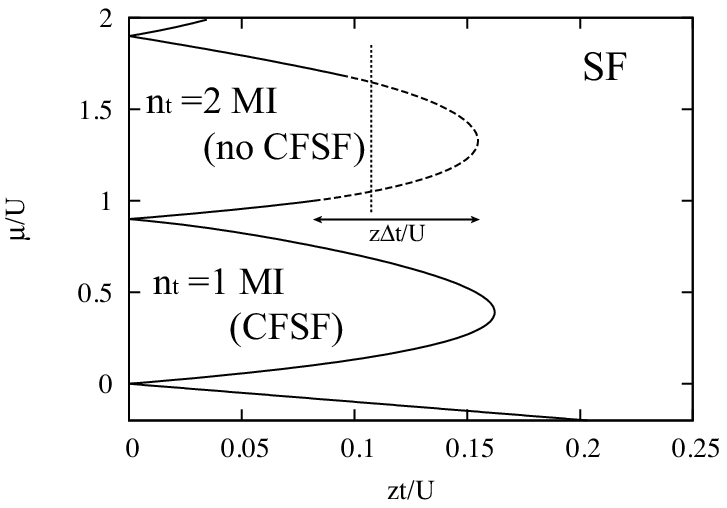}}%
\subfigure[]{\includegraphics[width=7.2cm]{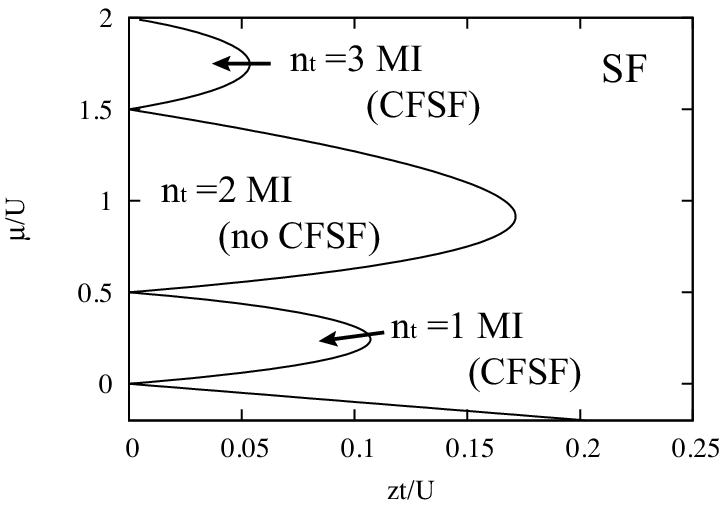}}%
\end{tabular}
\begin{tabular}{cc}
\subfigure[]{\includegraphics[width=7.2cm]{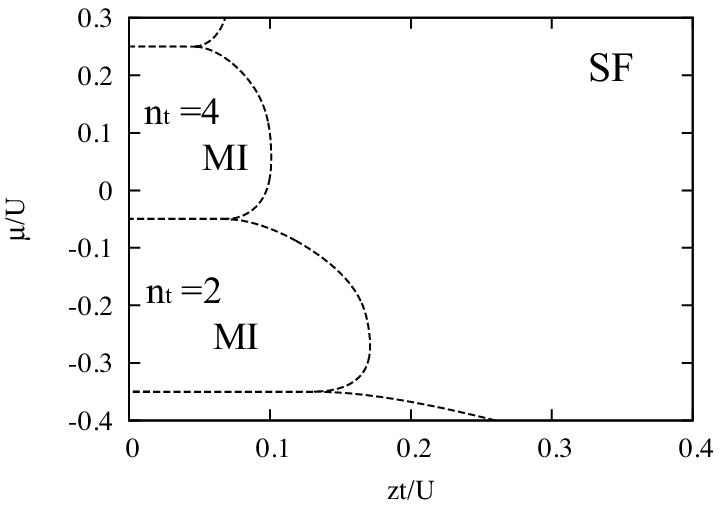}}%
\subfigure[]{\includegraphics[width=7.2cm]{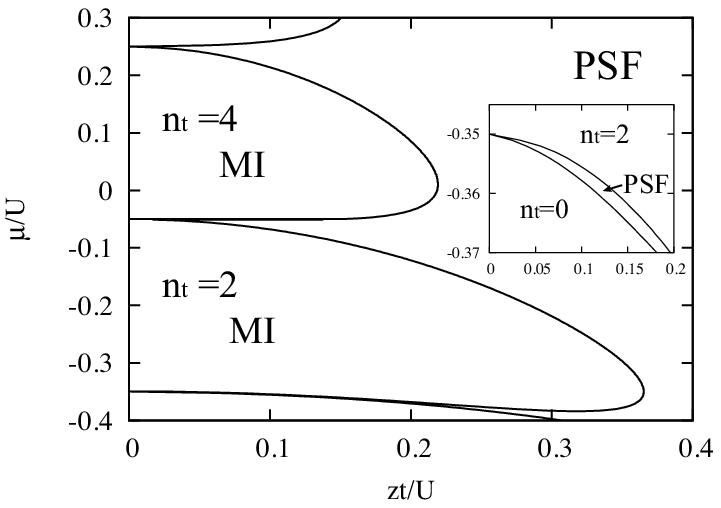}}%
\end{tabular}
\caption{\label{fig:diagram1} Phase diagrams on the $(zt/U, \mu/U)$ plane for (a) $U_{12}/U=0.9$, (b) $U_{12}/U=0.5$, and (c), (d) $U_{12}/U=-0.7$.
The solid lines and the dashed lines denote the phase boundaries for the second-order and the first-order transitions.
The phase boundaries in (a), (b), and (c) are obtained by solving Eq.~(\ref{eqn:GWe1}) while the phase boundaries between  PSF and MI in (d) are obtained by solving Eq.~(\ref{eqn:GWe2}).
The inset in (d) shows a magnified view that focuses on the PSF region between vacuum and the $n_{\rm t}=2$ MI.
In (a), $\Delta t$ represents the width of the region of first-order transition in the $t$-axis and the dotted line is the trajectory used in Fig.~\ref{fig:tot_den}.}
\end{center}
\end{figure}

In Figs.~\ref{fig:diagram1}(a) and (b), we depict the phase diagrams for repulsive inter-component interaction.
We see that there is a region of an incompressible MI for each integer value of the total filling factor.
The existence of the MI phases with odd fillings has been predicted in Refs.~\cite{PRL_90_100401, PRA_68_053602,  NJP_5_113}.
In the atomic limit ($t/U=0$), the size of the MI regions in the $\mu$-axis for odd and even total fillings is $U_{12}$ and $U$, respectively, and the odd-filling MI regions vanish at $U_{12}/U \rightarrow 0$~\cite{PRB_72_184507}.
Once the regions of the MI phases are located, we solve Eq.~(\ref{eqn:GWe3}) in order to check whether or not the CFSF order is present.
It is shown that the CFSF order emerges in the MI phases with odd total fillings  while it is not present in MI with even total fillings.
This result is consistent with that in Refs.~\cite{PRL_90_100401, NJP_5_113, PRL_92_050402, PRA_77_015602, PRA_80_023619, PRB_80_245109}.
Within the mean-field approximation used here, the local state of the ground state at odd total filling is $\frac{1}{\sqrt{2}}\left(|\frac{n_{\rm t} + 1}{2}, \frac{n_{\rm t} - 1}{2}\rangle+|\frac{n_{\rm t} - 1}{2}, \frac{n_{\rm t} + 1}{2}\rangle\right)$ while that at even total filling is $|\frac{n_{\rm t}}{2},\frac{n_{\rm t}}{2}\rangle$.
As indicated by the dashed line in Fig.~\ref{fig:diagram1}(a), when $U_{12}$ is close to $U$, there is a wide region of the first-order phase transition to MI with $n_{\rm t} = 2$ on the phase boundary.
We note that similar first-order transitions have been found in multi-component Bose-Hubbard systems with inter-component exchange interactions~\cite{PRL_94_110403, PRA_70_063601, PRA_71_033623, JPSJ_75_074601, PRB_84_064529}.
In Sec.~\ref{subsec:FOP}, we discuss the first-order transitions in detail.

In Figs.~\ref{fig:diagram1}(c) and (d), we show the phase diagrams for $U_{12}=-0.7$; the former is obtained by applying the GA directly to the original Hamiltonian of Eq.~(\ref{eqn:Ham}) while the latter is obtained from the effective Hamiltonian of Eq.~(\ref{eqn:efH1}).
In the phase diagrams, there are only the MI phases with even total fillings, and the local state of the MI phase is $|\frac{n_{\rm t}}{2}, \frac{n_{\rm t}}{2} \rangle$.
We see that the phase transition between SF and MI is entirely of the first order.
Although the first-order SF-to-MI transitions for attractive inter-component interaction have been previously found at commensurate fillings~\cite{PRL_92_050402} and in the hardcore limit~\cite{PRB_82_180510}, our result that extends the region of the first-order transition is complimentary to the previous findings.
In Fig.~\ref{fig:diagram1}(c), it is shown that there are direct transitions between different MI phases even when $t/U > 0$.
However, this is an artifact stemming from the fact that the approach fails to describe the PSF phase because of the lack of the second-order hopping process.
Indeed, when the phase diagram is calculated from the effective Hamiltonian as shown in Fig.~\ref{fig:diagram1}(d), the MI phases are separated from one another by thin but finite regions of PSF.
Notice that although the MI regions in Fig.~\ref{fig:diagram1}(d) are significantly larger than those in Fig.~\ref{fig:diagram1}(c), this happens because the effective Hamiltonian is invalid for a relatively large value of $zt/U \gtrsim 0.1$.
Thus, combining the information of the two phase diagrams, PSF is located in the regions of small hopping $zt/U \lesssim 0.1$ sandwiched between MI regions.

\subsection{\label{subsec:FOP}First-order phase transitions}

%
\begin{figure}[tb]%
\begin{center}
\begin{tabular}{cc}
\subfigure[]{\includegraphics[width=6.9cm,clip]{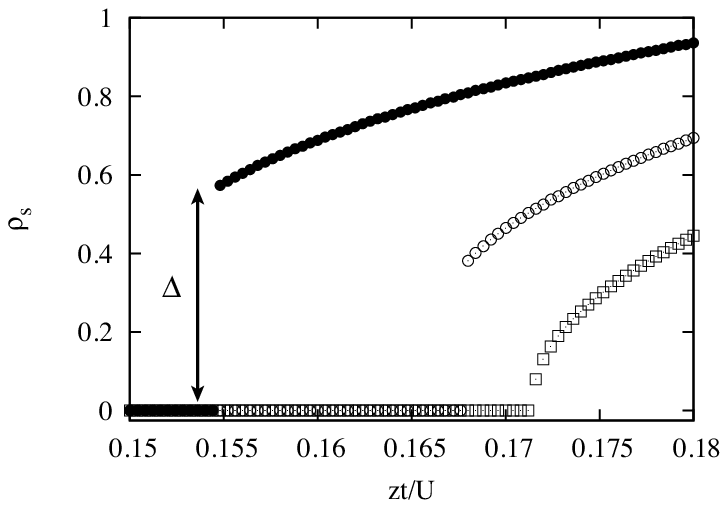}}%
\subfigure[]{\includegraphics[width=7.2cm,clip]{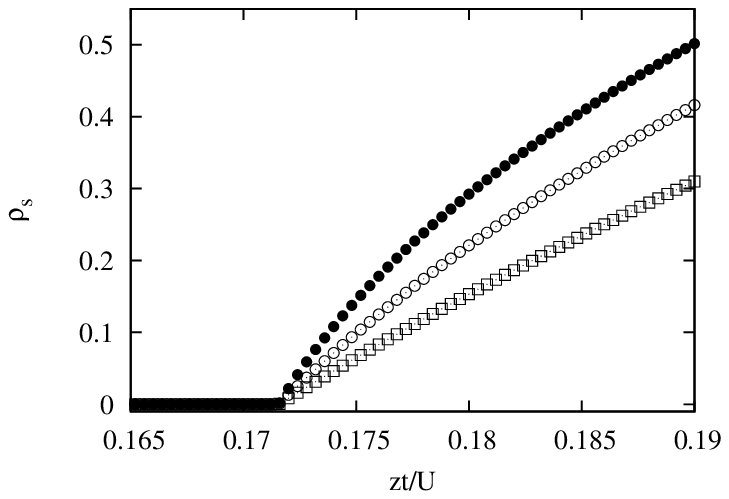}}%
\end{tabular}
\caption{\label{fig:trans} Condensate density $\rho_s=|\Phi_1|^2+|\Phi_2|^2$ with $n_{\rm{t}}=2$.
(a): The filled circles, the open circles, and the open squares denote data for $U_{12}/U=0.9$, $0.8$, and  $0.7$.
(b): The filled circles, the open circles, and the open squares denote data for  $U_{12}/U=0.6$, $0.5$, and $0.1$.}
\end{center}
\end{figure}
Let us now focus on the first-order phase transitions from SF to MI with even total fillings for repulsive inter-component interaction.
To confirm the first-order transition at commensurate fillings, which has been pointed out for the effective two-component $J$-current model in Ref.~\cite{PRL_92_050402}, we show in Fig.~\ref{fig:trans} the condensate density $\rho_s\equiv|\Phi_1|^2+|\Phi_2|^2$ versus $zt/U$ across the transition point for $n_{\rm t} = 2$ and several values of $U_{12}/U$.
In Fig.~\ref{fig:trans}(a), we plot $\rho_s$ for the values of $U_{12}/U$ at which transition is of the first order.
The transition point is determined as the crossing point of the energy of the MI state and that of the SF state.
At the transition point, the condensate density exhibits a jump, which we define as $\Delta$, and the magnitude of the jump depends on $U_{12}/U$. We find that the transition point $(zt/U)_{\rm tr}$ decreases with increasing $U_{12}/U$.
Notice that although we show in Fig.~\ref{fig:trans}(a) only the case of $n_{\rm t}=2$, we have checked that the presence of the first-order SF-to-MI transitions is a general feature in even total fillings.
In Fig.~\ref{fig:trans}(b), we show $\rho_s$ for the values of $U_{12}/U$ at which the transition is of the second order.
In this case, $\rho_s$ changes continuously across the transition point  and the phase transition occurs at $zt/U = (zt/U)_{\rm tr} = 0.172$ regardless the value of $U_{12}/U$.

\begin{figure}[tb]%
\begin{center}
\begin{tabular}{cc}
\subfigure[]{\includegraphics[width=7.2cm,clip]{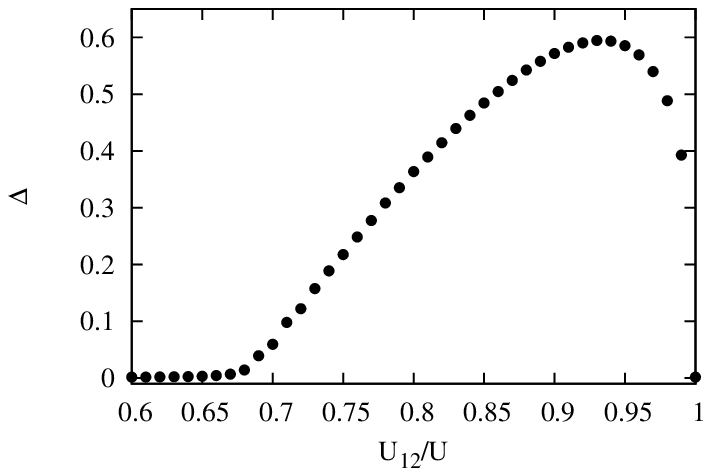}}%
\subfigure[]{\includegraphics[width=7.2cm,clip]{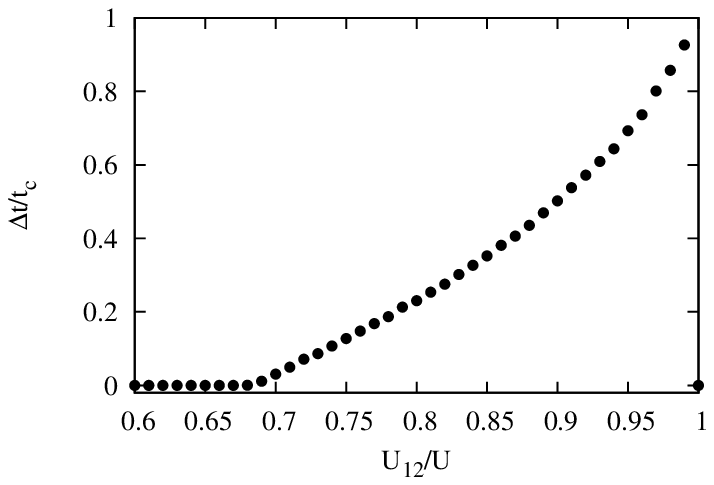}}%
\end{tabular}
\caption{\label{fig:gap} (a) Jump of the condensate density at the transition point $\Delta$ as a function of $U_{12}/U$ at the top of the $n_{\rm t}=2$ Mott lobe.
(b) The width of the region of first-order transition in the $t$-axis $\Delta t$, which is indicated in Fig.~\ref{fig:diagram1}(a) is shown in the unit of $t_{\rm c}$ as a function of $U_{12}/U$.}
\end{center}
\end{figure}

In order to determine the parameter regions where the transition is of the first order, we plot in Fig.~\ref{fig:gap}(a) the jump $\Delta$ as a function of $U_{12}/U$ at the top of the $n_{\rm t}=2$ Mott lobe.
$\Delta$ is zero for $U_{12}/U\lesssim 0.65$ while it is finite, i.e. the transition is of the first order, for $U_{12}/U\gtrsim 0.65$.
$\Delta$ is peaked at $U_{12}/U \simeq 0.93$, and it reaches zero at $U_{12}/U=1$.
Recall that the phase separation occurs when $U_{12}/U >1$, and the SF-to-MI transitions in the phase separated gases are of the second order, because they are equivalent to those of one-component bosons.

We next consider the transition between SF with incommensurate fillings and MI.
In Fig.~\ref{fig:gap}(b), we plot the width of the first-order transition region in the $t$-axis $\Delta t / t_c$ as a function of $U_{12}/U$, where $t_c$ is the hopping for the transition at the top of $n_{\rm t} = 2$ MI lobe.
We find that $\Delta t / t_c$ increases with increasing $U_{12}/U$.
In the limit of $U_{12}/U\to1$, $\Delta t / t_c$ reaches unity; this means that the entire phase boundary becomes of the first order.
However, $\Delta t / t_c$ suddenly drops to zero at $U_{12}/U = 1$ because the transition becomes of the second order.

\begin{figure}[hts]%
\begin{center}
\includegraphics[width=7.2cm]{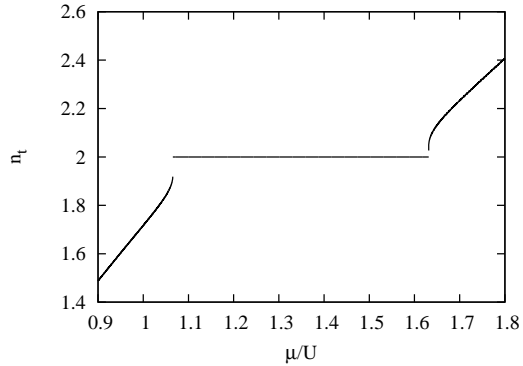}%
\caption{\label{fig:tot_den} Total density $n_{\rm t}$ as a function of $\mu/U$ along $zt/U=0.112$ line for $U_{12}/U=0.9$. The trajectory is indicated by the dotted line in Fig.~\ref{fig:diagram1} (a).}
\end{center}
\end{figure}

We plot the total filling factor $n_{\rm t}$ as a function of $\mu/U$ along $zt/U=0.112$ line for $U_{12}/U=0.9$ in Fig.~\ref{fig:tot_den}.
We see that the total density exhibits a jump at the transition points, which is similar to the condensate density jump discussed above. 
For $zt/U=0.112$, it is found that the density jumps are 0.082 at $\mu/U\simeq1.07$ and 0.02 at $\mu/U \simeq 1.63$.
For a fixed value of $zt/U$ the density jump at the small chemical potential is always larger than that at the large chemical potential.
It is worth noting that measuring the density jump is a possible option to observe the first order transition in experiment.
In the presence of a trapping potential, the local chemical potential spatially varies such that the transition from SF with commensurate filling to MI can occur.
Within the local density approximation, the density jumps emerge at the edges of the $n_{\rm{t}}=2$ MI plateau.
Given the fact that first order transitions in a system of Fermi-Fermi mixtures with population imbalance have been observed by measuring density jumps~\cite{Nature_451_689}, it is expected that this way may work also for the first-order SF-to-MI transitions in Bose-Bose mixtures discussed here.

\section{\label{sec4}Excitation spectra}
In Sec.~\ref{sec2}, we presented the formulation for calculating elementary excitations of a Bose-Bose mixture in a $D$-dimensional hypercubic optical lattice.
Having obtained the ground-state phase diagrams in Sec.~\ref{sec3}, we use the formulation to reveal properties of excitation spectra in this section.
Henceforth, we assume $D=2$ and the momentum of excitations to be $k_x=k_y \equiv k$.

\subsection{\label{subsec:Exc_MI}MI phase}

We first consider the MI phase with even total fillings and analytically calculate the excitation spectrum.
For the MI phase of this type, the variational parameters are given by $\tilde{f}_{n_1,n_2}=\delta_{n_1,n}\delta_{n_2,n}$ and the superfluid order parameters $\Phi_{a}$ are equal to zero.
Hence, the linearized equations of motion (\ref{eqn:GW-bogo_1}) take $8\times8$ matrix form.
Substituting these parameters into Eq.~(\ref{eqn:GW-bogo_1}) and solving it, we obtain the excitation energy,
\begin{align}
\label{eqn:exc_MI1}
\hbar\omega_{\bf k}^{(\pm)}&=\frac{1}{2}\left[ \sqrt{ \epsilon(\mathbf{k})^2-\epsilon(\mathbf{k}) U\left(4n+2\right)+U^2} \pm\left(\epsilon(\mathbf{k})-U(2n-1)-2U_{12}+2\mu \right) \right],
\end{align}
where the plus (minus) sign corresponds to the particle (hole) excitation.
The particle and hole excitations of component $1$ are degenerate with those of component $2$, and this means that there are four excitation branches in total.
There are other solutions of Eq.~(\ref{eqn:GW-bogo_1}) that are independent of ${\bf k}$ with the energies,
\begin{align}
\label{eqn:exc_MI2}
\hbar\omega_{m_1,m_2}=&\ \frac{U}{2}m_1(m_1-1)+\frac{U}{2}m_2(m_2-1)+U_{12}m_1m_2-\mu m_1-\mu m_2\notag\\
&-\frac{U}{2}n_1(n_1-1)-\frac{U}{2}n_2(n_2-1)-U_{12}n_1n_2 +\mu n_1+\mu n_2,
\end{align}
where $m_{\alpha}$ is an non-negative integer other than $(m_1,m_2) = (n_{1}, n_{2}\pm1)$ or $(n_{1}\pm1, n_{2})$.
Each branch corresponds to multi-particle or multi-hole excitation and its energy is always positive.

\begin{figure}[hts]%
\centering
\includegraphics[width=7.2cm]{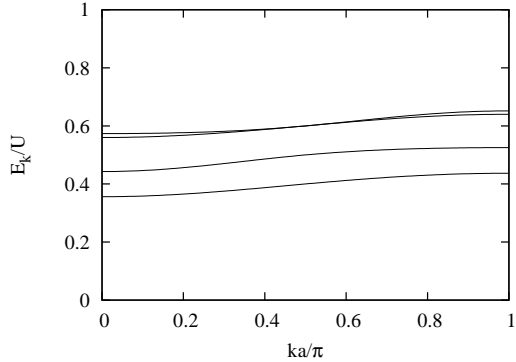}%
\caption{\label{Fig:exc_mi_p}Excitation spectrum in the MI phase with $n_{\rm{t}}=1$, $U_{12}/U=0.9$, $t/U=0.01$, and $\mu/U=0.4$.
The lowest branch is a hole excitation and the upper branches are particle excitations.
}
\end{figure}

We next consider the MI phase with odd total fillings.
In this phase, the variational parameters are given by $f^{(i)}_{n_1,n_2}=\frac{1}{\sqrt{2}}\left(\delta_{n_1,(n_{\rm t}+1)/2}\delta_{n_2,(n_{\rm t}-1)/2}+\delta_{n_1,(n_{\rm t}-1)/2}\delta_{n_2,(n_{\rm t}+1)/2}\right)$.
Inserting these parameters into Eq.~(\ref{eqn:GW-bogo_1}), we obtain the excitation spectrum by diagonalizing the Bogoliubov equation.
For the case of the MI phase with $n_{\rm{t}}=1$, we obtain 
\begin{align}
\left( U-\mu-\epsilon(\mathbf{k})\pm\hbar\omega \right)\left[ \left( \mu\mp\hbar\omega \right)\left\{ \left( U_{12}-\mu-\epsilon(\mathbf{k})\pm\hbar\omega \right)\left( U-\mu-\epsilon(\mathbf{k})\pm\hbar\omega \right) -\epsilon(\mathbf{k})^2 \right\} \right.&\notag\\
\left.\hspace{3cm}-\epsilon(\mathbf{k})\left( U_{12}-\mu\pm\hbar\omega \right)\left( U-\mu\pm\hbar\omega \right) \right]&=0.
\end{align}
In Fig.~\ref{Fig:exc_mi_p}, we plot the excitation spectrum of the MI phase with $n_{\rm t}=1$.
There we see four branches in this MI phase, which correspond to one hole excitation and three particle excitations.
In contrast to the MI phase, the excitations are not degenerate.
There are other multi-particle excitations, which we do not explicitly show here.

%

\subsection{\label{subsec:Exc_SF}SF phase}
%
\begin{figure}[tb]%
\begin{center}
\begin{tabular}{cc}
\subfigure[]{\includegraphics[width=7.2cm]{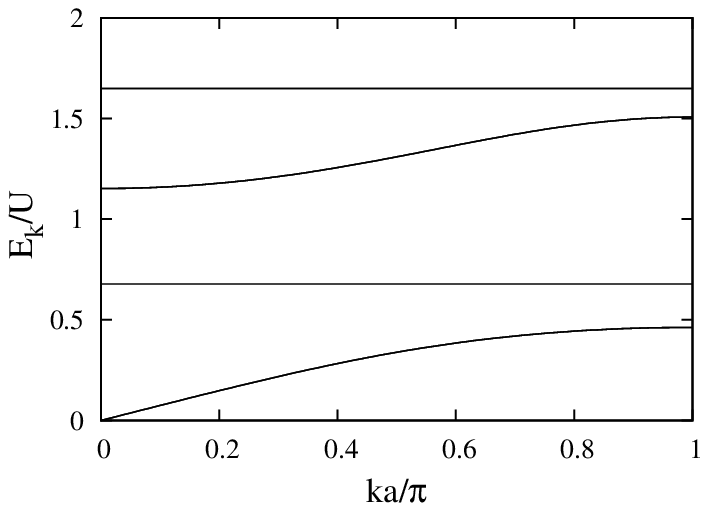}}%
\subfigure[]{\includegraphics[width=7.2cm]{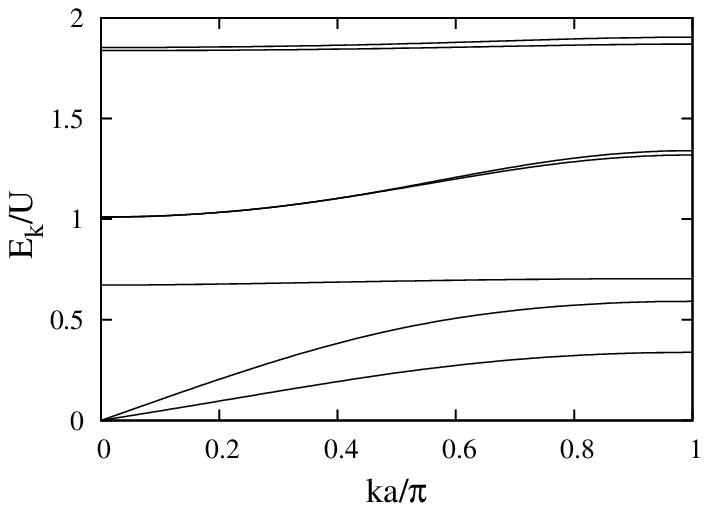}}%
\end{tabular}
\begin{tabular}{cc}
\subfigure[]{\includegraphics[width=7.2cm]{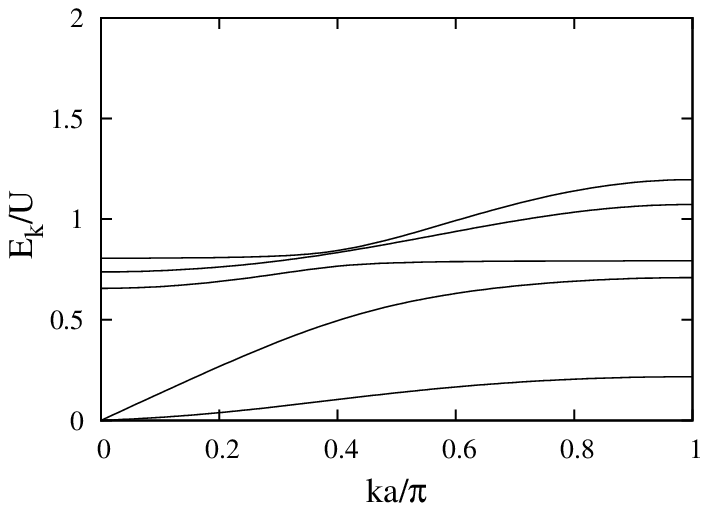}}%
\subfigure[]{\includegraphics[width=7.2cm]{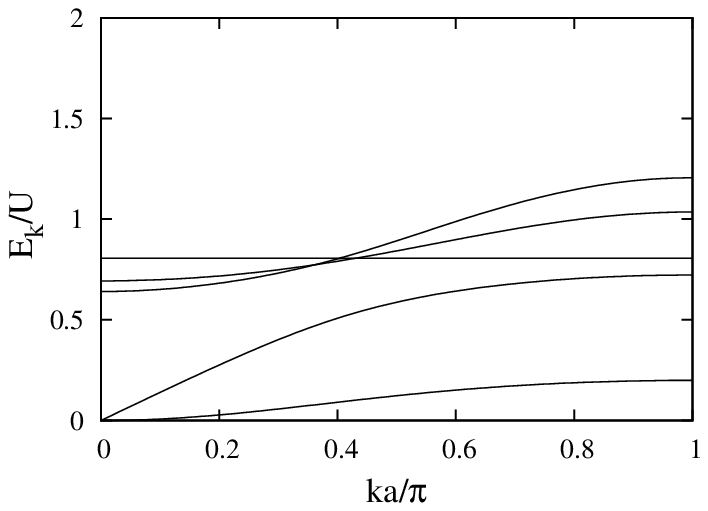}}%
\end{tabular}
\caption{\label{Fig:exc_sf_p} Excitation spectra of the SF phase with $n_{\rm t}=1$ at $t/U=0.07$ for $U_{12}\geq 0$.
We set inter-component interactions as $U_{12}/U=0$ (a), $0.4$ (b), $0.9$ (c), and $1$ (d), respectively.}
\end{center}
\end{figure}

We now calculate the excitation spectra in the SF phase by numerically diagonalizing Eq.~(\ref{eqn:GW-bogo_1}).
In Fig.~\ref{Fig:exc_sf_p}(a), we plot the excitation spectrum for $n_{\rm t}=1$ and $U_{12}=0$.
There we see two dispersive modes and two non-dispersive modes.
While the gapful dispersive mode near the MI transition corresponds to oscillations of the amplitude of the order parameters, the gapless dispersive mode is a phase-fluctuation mode called the Bogoliubov spectrum~\cite{PRA_76_063607}.
Since the system consists of two independent and equivalent one-component Bose gases, the dispersive modes agree with those for a one-component Bose gas~\cite{PRA_84_033602} and each branch is doubly degenerate.
In contrast, the non-dispersive modes are particular to the Bose-Bose mixture.

We plot the excitation spectra for the repulsive inter-component interaction in Figs.~\ref{Fig:exc_sf_p}(b)-(d).
It is clearly seen that finite $U_{12}$ splits degenerate modes into in-phase and out-of-phase modes.
Increasing $U_{12}$, the two gapless modes repel each other more strongly than the amplitude modes.
In Fig.~\ref{Fig:exc_sf_p}(c), we find that a level repulsion occurs around $ka/\pi=0.4$ and the mode that was non-dispersive at $U_{12}=0$ acquires $k$-dependence.
When $U_{12}/U=1$, a non-dispersive mode appears again in Fig.~\ref{Fig:exc_sf_p}(d).
Furthermore, for $U_{12}/U>1$, the gapless out-of-phase mode has an imaginary part around $k=0$, which means the dynamical instability toward phase separation.

\begin{figure}[tb]%
\centering
\begin{tabular}{cc}
\subfigure[]{\includegraphics[width=7.2cm]{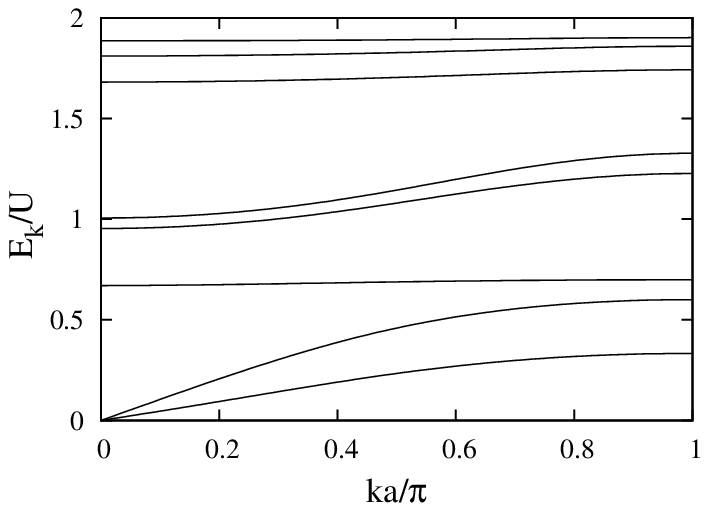}}%
\subfigure[]{\includegraphics[width=7.2cm]{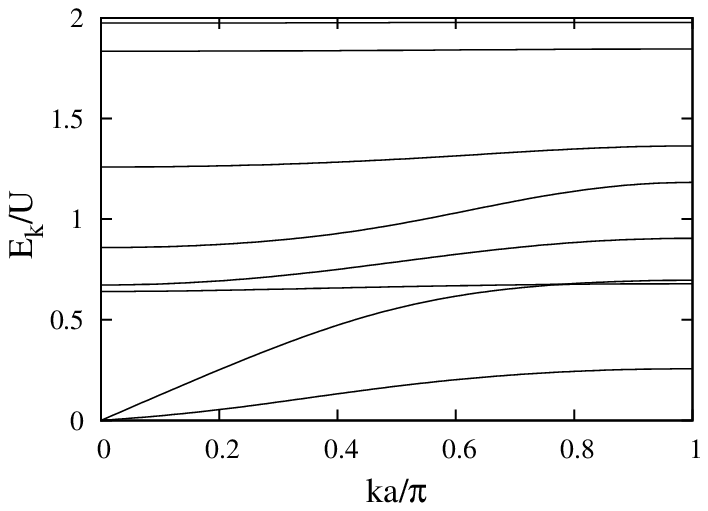}}%
\end{tabular}
\caption{\label{Fig:exc_sf_m}Excitation spectra of the SF phase with $n_{\rm{t}}=1$ at $t/U=0.07$, $U_{12}=-0.4$ (a) and $-0.7$ (b).}
\end{figure}

In Fig.~\ref{Fig:exc_sf_m}, we plot the excitation spectra for $U_{12}<0$.
We again see the repulsion of gapless modes and the shift of other modes compared with that of $U_{12}=0$.
One of the main differences from the case of $U_{12}>0$ is that the gapless in-phase mode exhibits dynamical instability, which leads to collapse of the mixture, for strong attractive interaction.
In our calculations, the mixture collapses at $\mu/U=-0.79$ for $t/U=0.07$ and $n_{\rm{t}}=1$.

\subsection{\label{subsec:Exc_PT}Around the phase transition}
%
\begin{figure}[tb]%
\begin{center}
\begin{tabular}{ccc}
\subfigure[]{\includegraphics[width=5.4cm]{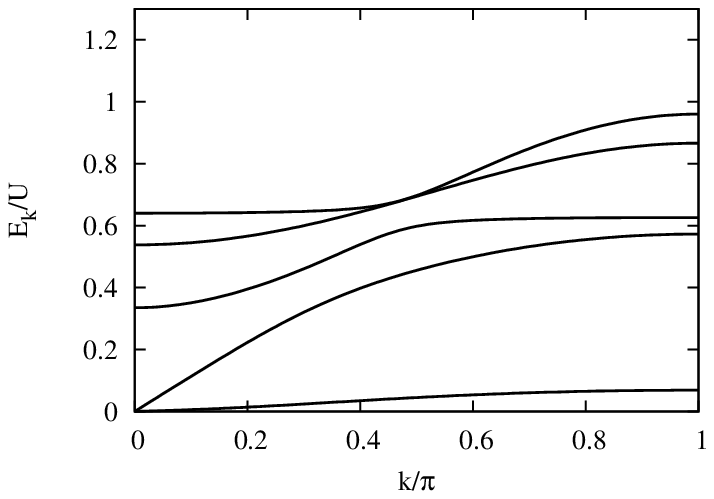}}%
\subfigure[]{\includegraphics[width=5.4cm]{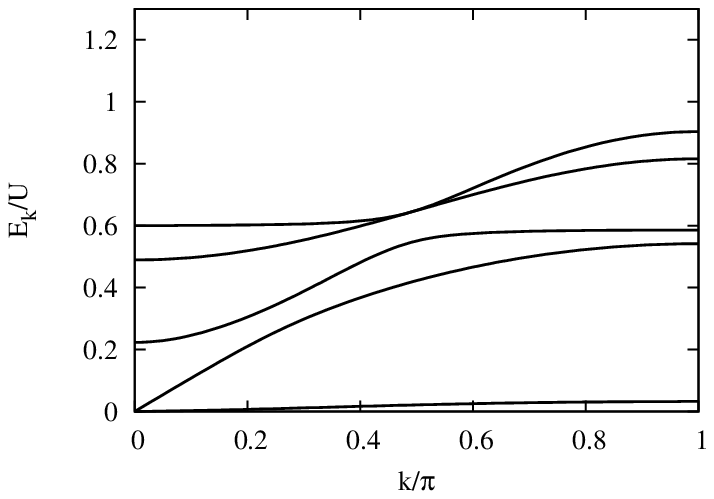}}%
\subfigure[]{\includegraphics[width=5.4cm]{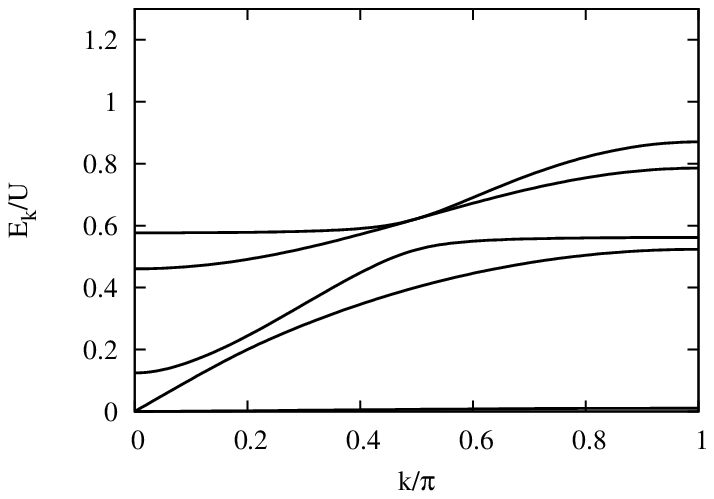}}%
\end{tabular}
\begin{tabular}{ccc}
\subfigure[]{\includegraphics[width=5.4cm]{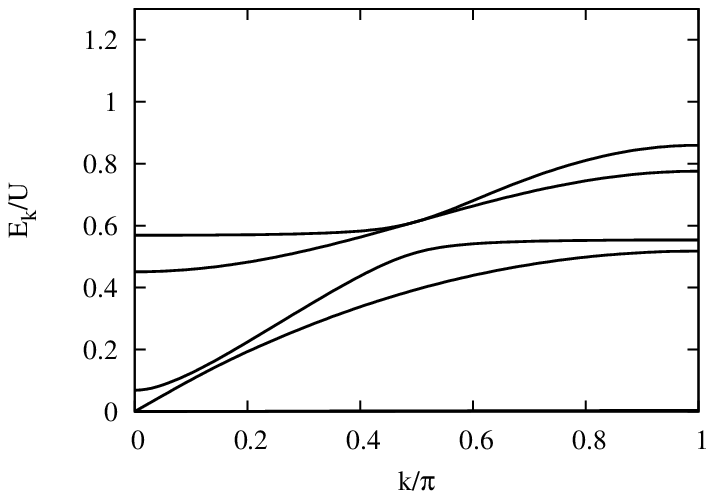}}%
\subfigure[]{\includegraphics[width=5.4cm]{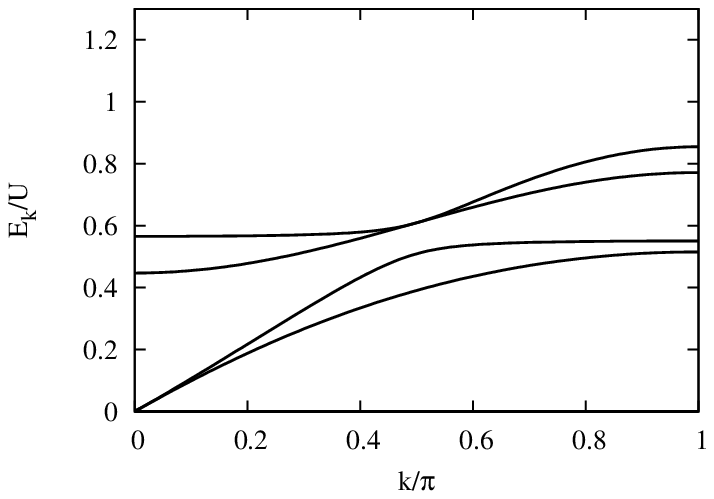}}%
\subfigure[]{\includegraphics[width=5.4cm]{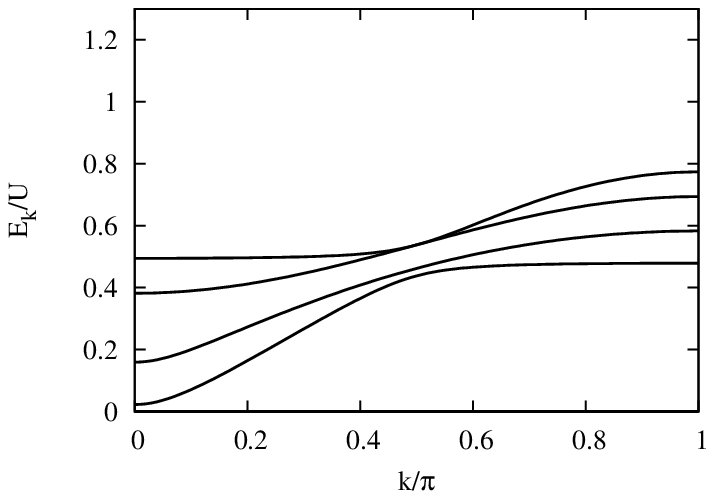}}%
\end{tabular}
\caption{\label{Fig:exc_u9_nt1}Excitation spectra along the $n_{\rm t}=1$ line at $U_{12}/U=0.9$.
The hopping amplitude is (a) $t/U=0.05$, (b) $t/U=0.045$, (c) $t/U=0.042$, (d) $t/U=0.041$, (e) $t/U=0.040557\simeq t_c/U$, and (f) $t/U=0.039$.}
\end{center}
\end{figure}
In Fig.~\ref{Fig:exc_u9_nt1}, we plot the excitation spectra for $n_{\rm t}=1$, $U_{12}/U=0.9$, and several values of $t/U$ across the SF-to-MI transition point  $t_c/U\simeq 0.0406$.
Notice that when we calculate the excitation spectrum of the MI phase, we use the same chemical potential as that at the transitional point.
When $t/U$ decreases from SF into MI, the following two remarkable changes happen in the excitation spectra.
One is that the in-phase amplitude (gapful) mode approaches the in-phase Bogoliubov (gapless) mode and that the amplitude mode becomes gapless at the transition point so that its low-energy part coincides with that of the Bogoliubov mode.
In the MI phase, the two branches split into gapful modes, corresponding to the particle and hole excitations.
This behavior is analogous to the case of one-component Bose-Hubbard model~\cite{PRL_89_250404, PRB_75_085106, PRA_84_033602, PRB_84_174522, PRL_109_010401}.
The other is that the out-of-phase mode becomes non-dispersive and reaches zero at the transition point.
It remains non-dispersive and zero in the MI phase.
However, this property is an artifact of our approach in which the second-order hopping process is neglected~\cite{NJP_5_113}.
In Sec.~\ref{subsec:Exc_PandC}, we show from calculations on the basis of the effective Hamiltonian that this mode has phonon-like dispersion, reflecting the counterflow superfluidity.

\begin{figure}[tb]%
\begin{center}
\begin{tabular}{ccc}
\subfigure[]{\includegraphics[width=5.4cm]{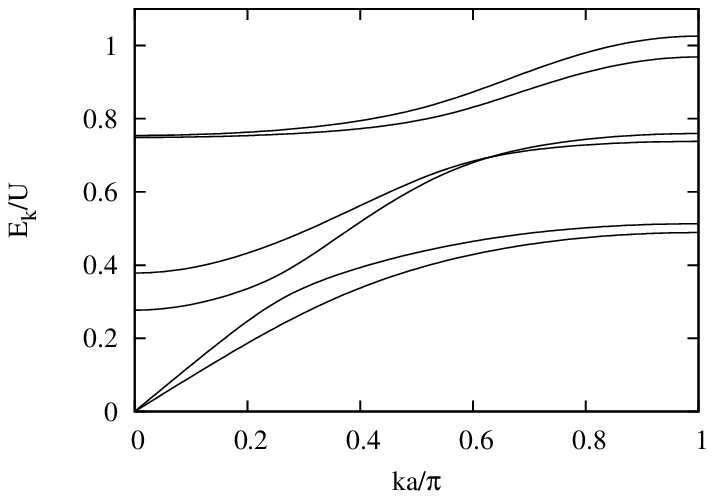}}%
\subfigure[]{\includegraphics[width=5.4cm]{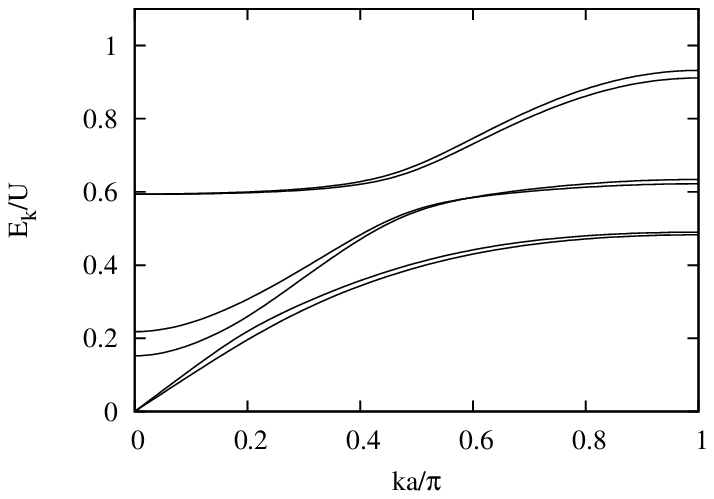}}%
\subfigure[]{\includegraphics[width=5.4cm]{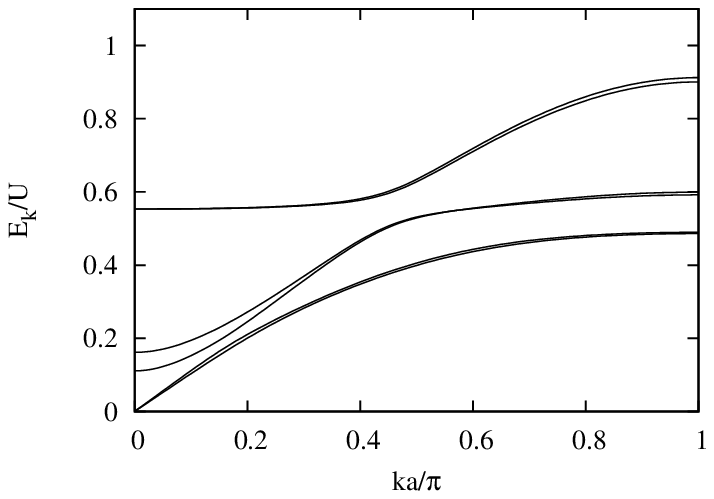}}%
\end{tabular}
\begin{tabular}{ccc}
\subfigure[]{\includegraphics[width=5.4cm]{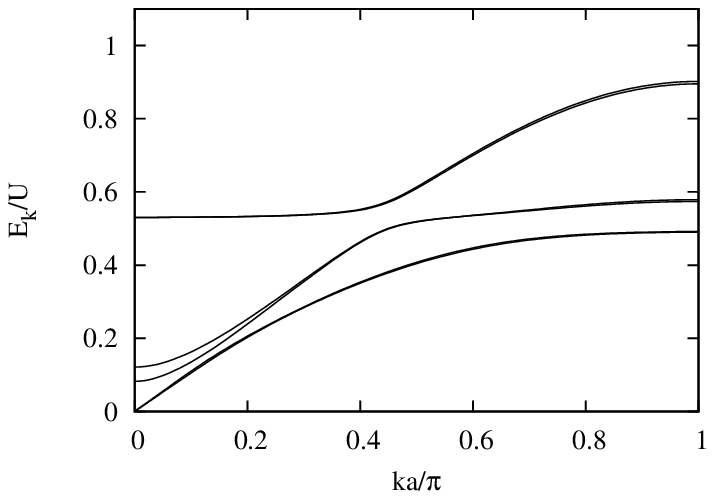}}%
\subfigure[]{\includegraphics[width=5.4cm]{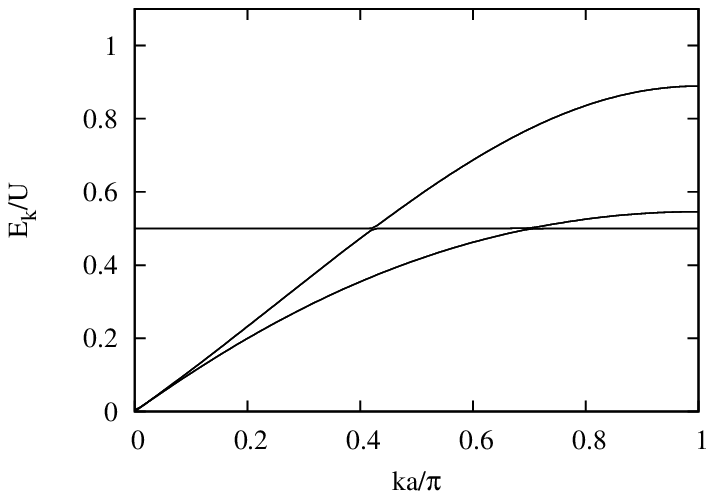}}%
\subfigure[]{\includegraphics[width=5.4cm]{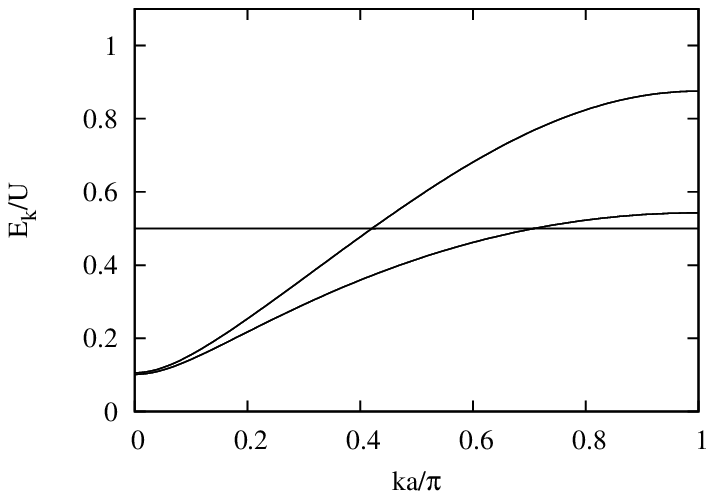}}%
\end{tabular}
\caption{\label{Fig:exc_u5_nt2}Excitation spectra along the $n_{\rm{t}}=2$ line at $U_{12}/U=0.5$.
The hopping amplitude is (a) $t/U=0.05$, (b) $t/U=0.045$, (c) $t/U=0.044$, (d) $t/U=0.0435$, (e) $t/U=0.04289\simeq t_c/U$, and (f) $t/U=0.041$.}
\end{center}
\end{figure}
In Fig.~\ref{Fig:exc_u5_nt2}, we plot the excitation spectra for $n_{\rm t}=2$, $U_{12}/U=0.5$, and several values of $t/U$.
When $t/U$ decreases in the SF phase, two of the gapful modes approach and their gaps descend toward zero.
Moreover, the two gapless modes also approach.
At the transition point, the low-energy parts of these four modes coincide.
This reflects the fact that both of the in-phase and out-of-phase superfluids disappear at the same time at the transition point.
In the Mott phase, there are a non-dispersive branch and two dispersive branches, which were discussed in Sec.~\ref{subsec:Exc_MI}.

\begin{figure}[tb]%
\begin{center}
\begin{tabular}{ccc}
\subfigure[]{\includegraphics[width=5.4cm]{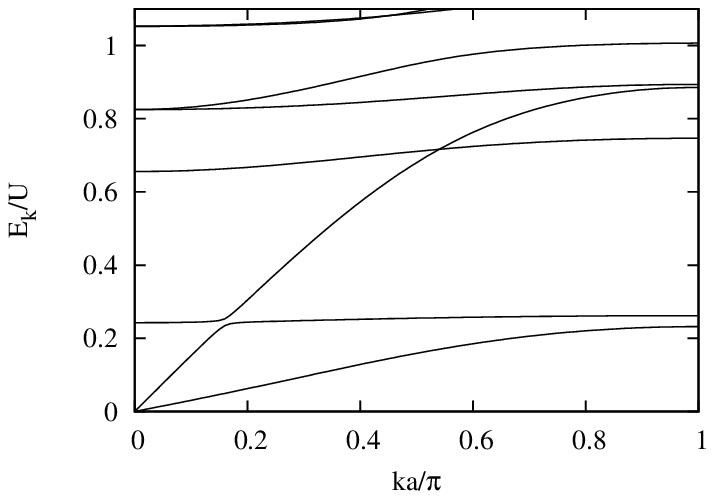}}%
\subfigure[]{\includegraphics[width=5.4cm]{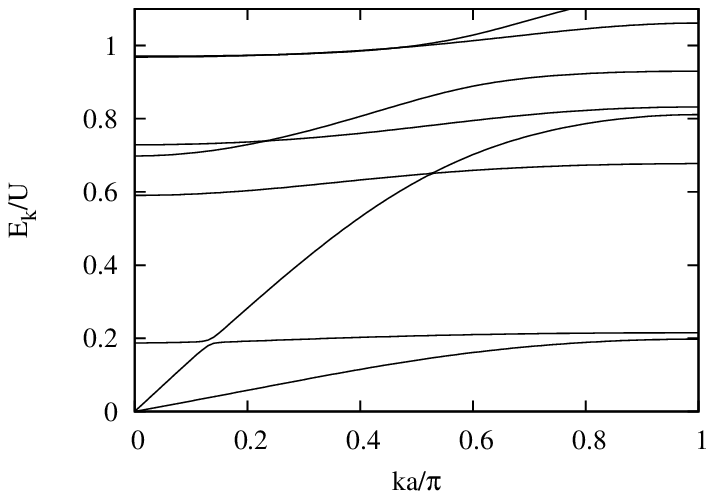}}%
\subfigure[]{\includegraphics[width=5.4cm]{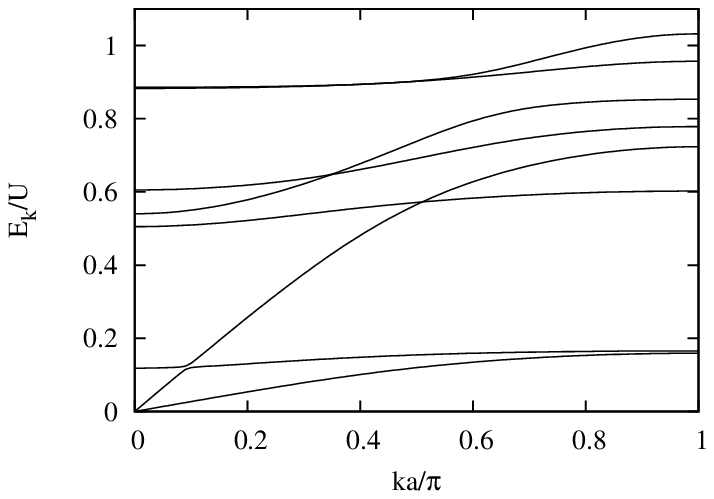}}%
\end{tabular}
\begin{tabular}{ccc}
\subfigure[]{\includegraphics[width=5.4cm]{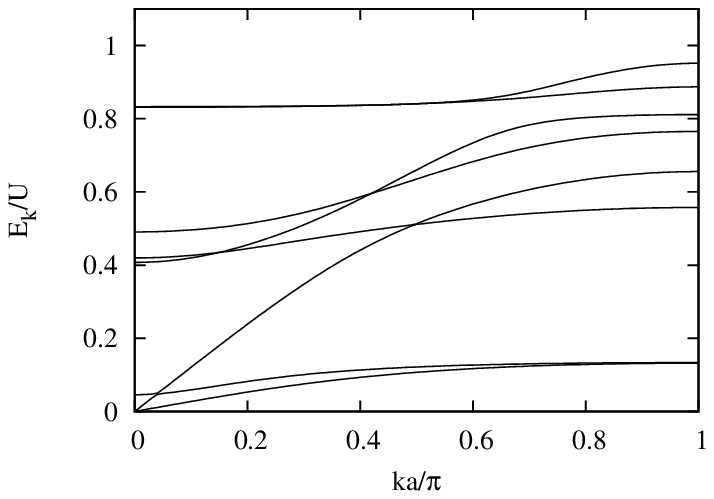}}%
\subfigure[]{\includegraphics[width=5.4cm]{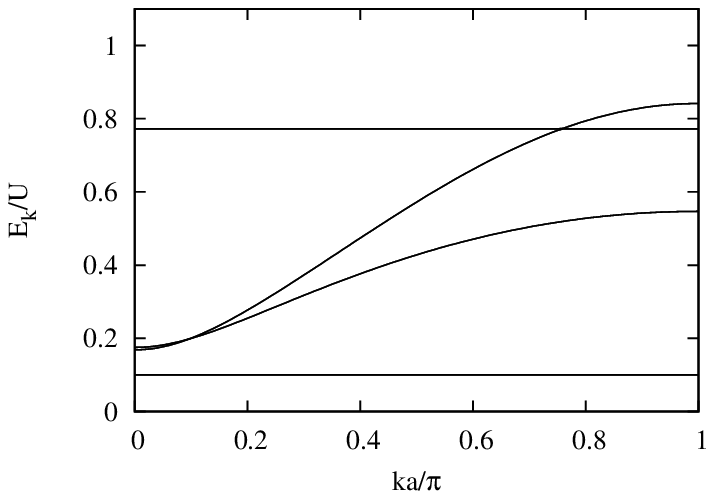}}%
\subfigure[]{\includegraphics[width=5.4cm]{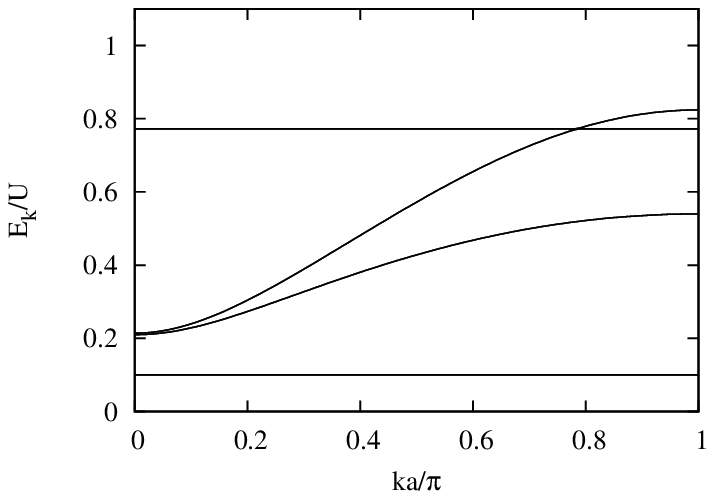}}%
\end{tabular}
\caption{\label{Fig:exc_u9_nt2}Excitation spectra along the $n_{\rm t}=2$ line at $U_{12}/U=0.9$.
The hopping amplitude is (a) $t/U=0.05$, (b) $t/U=0.045$, (c) $t/U=0.04$, (d) $t/U=0.0378$, (e) $t/U=0.0377$, and (f) $t/U=0.035$.}
\end{center}
\end{figure}

In Fig.~\ref{Fig:exc_u9_nt2}, we plot the excitation spectra for $n_{\rm t}=2$, $U_{12}/U=0.9$, and several values of $t/U$ across the SF-to-MI transition.
Because of the first-order transition, the excitation spectrum changes discontinuously at the transition point as clearly shown in Figs.~\ref{Fig:exc_u9_nt2}(d) and (e).
In contrast to the cases of the former two cases, any gapful modes do not coincide with the sound modes at the transition.
In the MI phase, there is a non-dispersive mode expressed by Eq.~(\ref{eqn:exc_MI2}) with $(m_1,m_2) = (2,0)$ or $(0,2)$ below the particle and hole modes.
This mode becomes dispersive when the second-order hopping processes are taken into account.

\subsection{\label{subsec:Exc_PandC}PSF and CFSF phase}
\begin{figure}[tb]%
\centering
\begin{tabular}{cc}
\subfigure[]{\includegraphics[width=7.2cm]{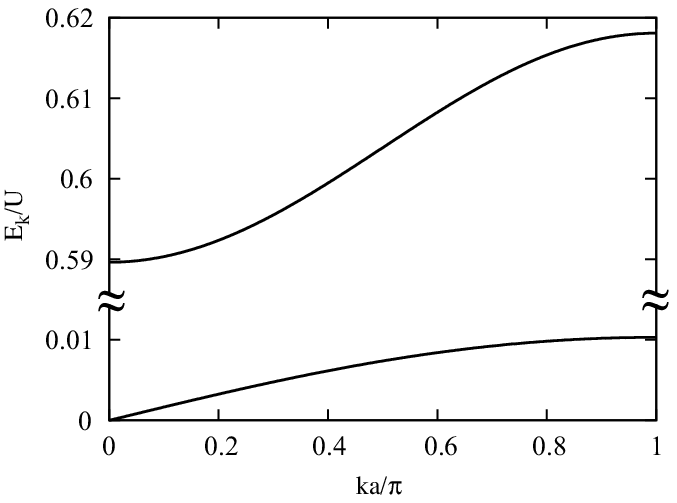}}%
\subfigure[]{\includegraphics[width=7.2cm]{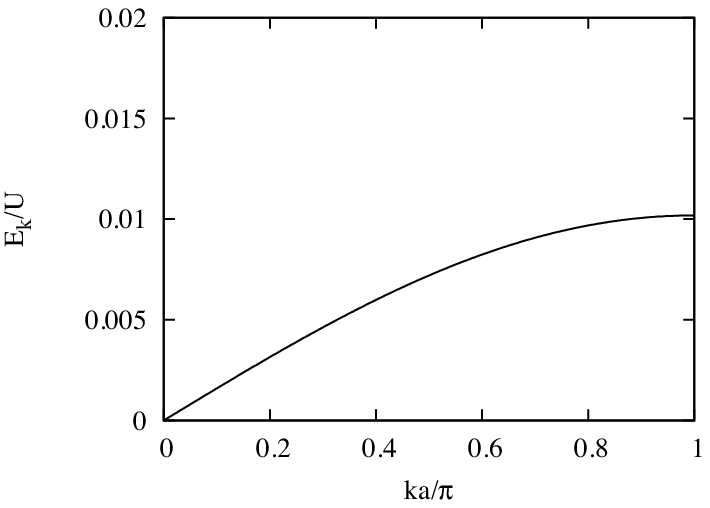}}
\end{tabular}
\caption{\label{Fig:exc_eff}Excitation spectra of the (a) PSF phase with $t/U=0.03$ and $U_{12}/U=0.9$, and (b) CFSF phase with $t/U=0.03$ and $U_{12}/U=-0.7$.}
\end{figure}

In Fig.~\ref{Fig:exc_eff}, we plot the excitation spectra in the PSF and CFSF phases with $n_{\rm t}=1$.
We set the parameters $t/U=0.03$ and $U_{12}/U=-0.7$ for the PSF phase and $t/U=0.03$ and $U_{12}/U=0.9$ for the CFSF phase.
While we see a gapless mode and a gapful mode in the PSF phase, only a gapless mode is present in the CFSF phase.
Gapful modes are absent in CFSF because the local state consists of only two Fock states $|1,0\rangle$ and $|0,1\rangle$.
It is well known that the local states have to consist of three or more Fock states in order for a gapful mode to appear in the excitation spectrum~\cite{PRL_89_250404,PRB_75_085106}.
If one calculates the excitation spectrum of the CFSF phase with $n_{\rm t} \ge 3$, one obtains gapful branches in addition to a gapless branch.
The gapless mode in each phase exhibits linear dispersion, which reflects superfluidity of pairs or anti-pairs.
As was discussed in Sec.~\ref{subsec:Exc_PT}, if we calculate the excitation spectrum for the same parameters by solving the linearized equations of motion (\ref{eqn:GW-bogo_1}) with the matrix elements of Eqs.~(\ref{eqn:matA}) and (\ref{eqn:matB}), the gapless modes are non-dispersive so that the Landau critical velocities are zero.
This means that PSF and CFSF acquire superfluidity through the second-order hopping process.

%

\section{\label{sec5}Dynamical structure factors}
Dynamical structure factors of one-component Bose gases in optical lattices have been experimentally measured by using the Bragg spectroscopy techniques, and this measurement has led to the observation of the Bogoliubov mode in the SF phase~\cite{Nature_Phys_6_56}.
In this section, we calculate the dynamical structure factors of Bose-Bose mixtures for SF, PSF, and CFSF phases to show that these phases can be distinguished through the Bragg spectroscopy.

In the Bragg spectroscopy, one exposes a Bose gas to an oscillating external field with momentum ${\bf k}$ and frequency $\omega$, which induces density fluctuations of the gas, and measures the response to the fluctuations.
Here we consider the following three types of density fluctuation with the form of Eq.~(\ref{eqn:a1}): (i) a density fluctuation to one-component $\hat{G}_{1, i} = \hat{F}_{1, i} = \hat{n}_{1, i}$, (ii) an in-phase density fluctuation $\hat{G}^{\rm in}_{i} = \hat{F}^{\rm in}_{i} = \hat{n}_{1, i} + \hat{n}_{2, i}$, and (iii) an out-of-phase density fluctuation $\hat{G}^{\rm out}_{i} = \hat{F}^{\rm out}_{i} = \hat{n}_{1,i}-\hat{n}_{2,i}$.
Substituting these fluctuations into the linear response formulae Eqs.~(\ref{eqn:RF}) and (\ref{eqn:LR_SQO}), we calculate the dynamical structure factors $S_{\hat{F}}({\bf k},\omega)$.
These types of density fluctuation can be implemented with use of component-dependent laser beams~\cite{PRL_105_045303, NJP_12_055013}.

\begin{figure}[hts]%
\centering
\begin{tabular}{c}
\subfigure[]{\includegraphics[width=7.2cm,clip]{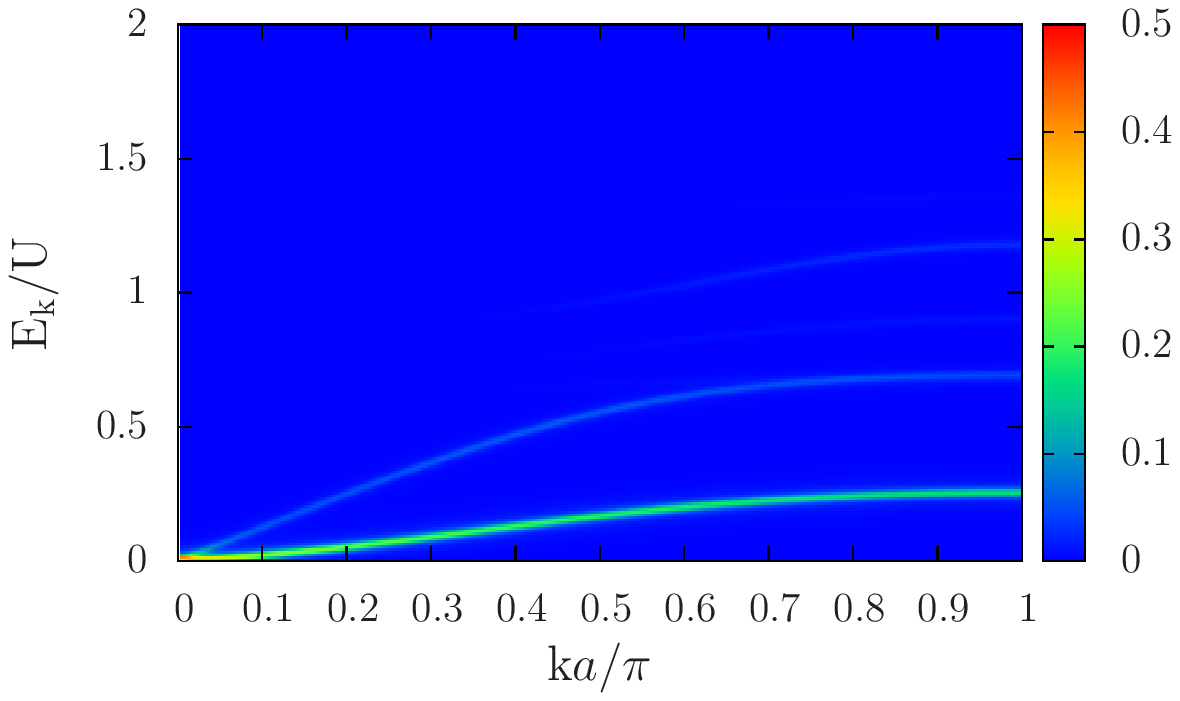}}
\end{tabular}
\begin{tabular}{cc}
\subfigure[]{\includegraphics[width=7.2cm,clip]{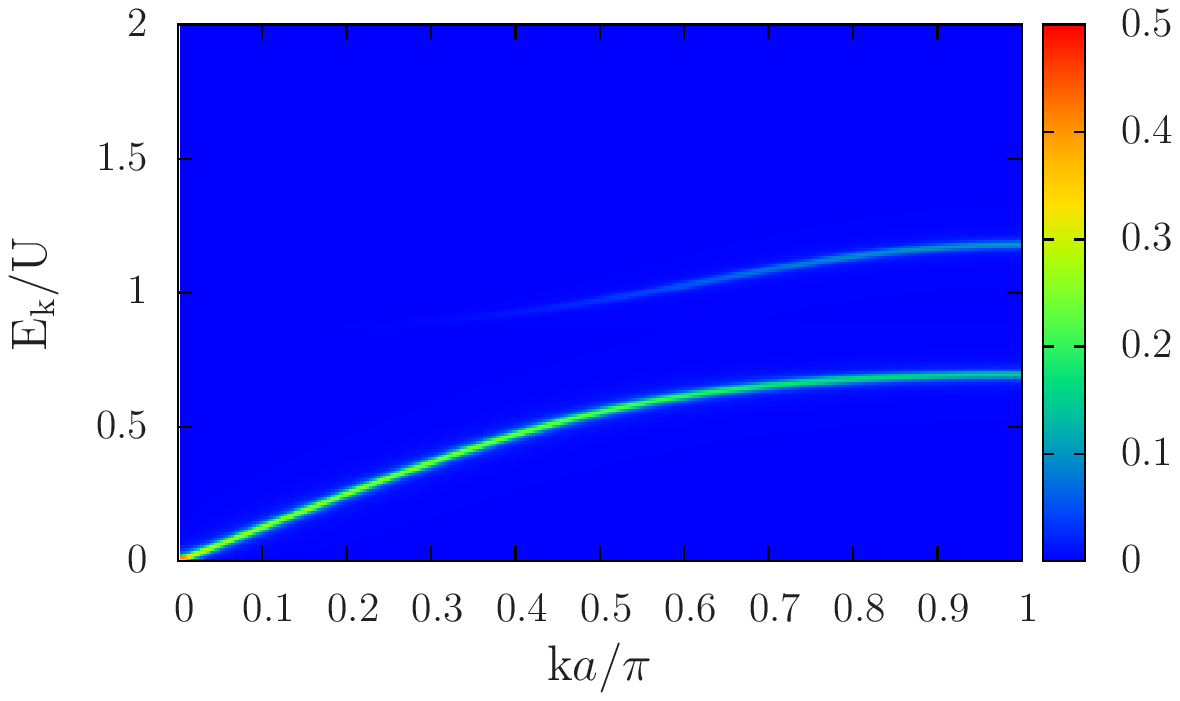}}
\subfigure[]{\includegraphics[width=7.2cm,clip]{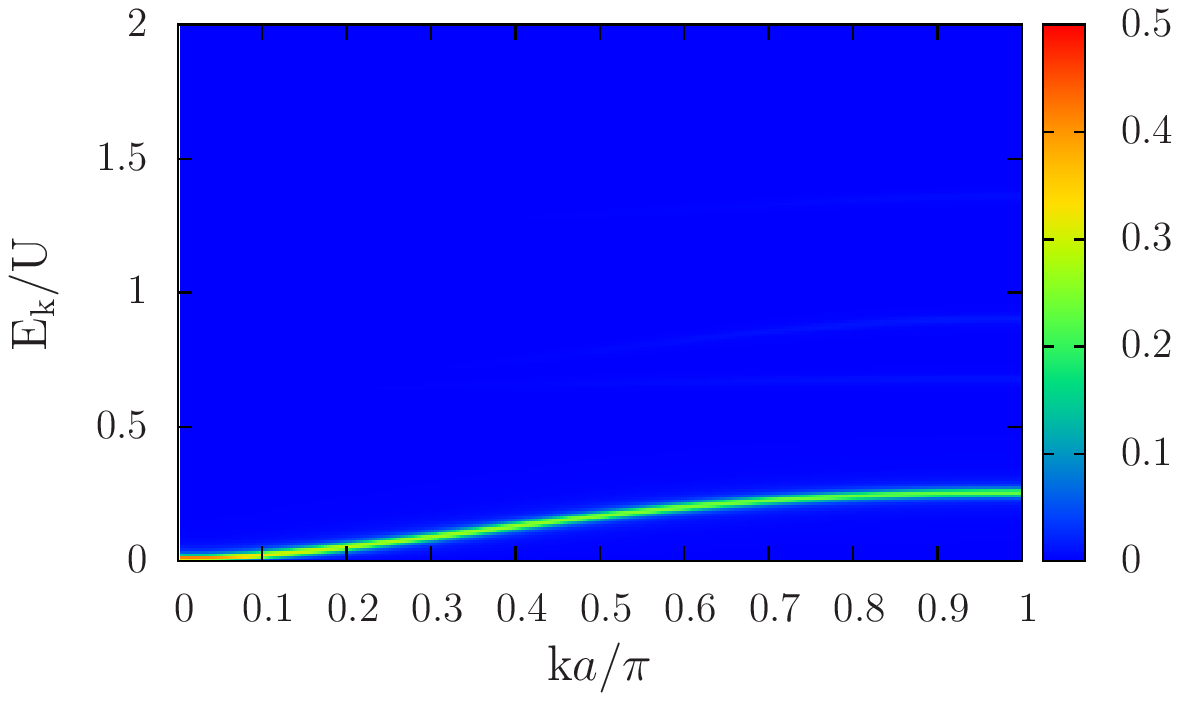}}
\end{tabular}
\caption{\label{Fig:DRF_sf_p}(Color online) Dynamical structure factors $\bar{S}_{\hat{F}}(\mathbf{k},\omega)$ of the SF phase with $n_{\rm t}=1$ for $t/U=0.07$ and $U_{12}/U=0.9$. Shown are the responses to the density fluctuation of one-component $\hat{F}_{1,i}$ (a), the in-phase density fluctuation $\hat{F}^{\rm in}_i$ (b), and the out-of-phase density fluctuation $\hat{F}^{\rm out}_i$ (c).}
\end{figure}

Let us first consider the SF phase.
In Fig.~\ref{Fig:DRF_sf_p}, we plot the dynamical structure factors for $n_{\rm t} = 1$, $t/U=0.07$, and $U_{12}/U=0.9$.
The parameters are the same as those in Fig.~\ref{Fig:exc_sf_p}(c).
As shown in Fig.~\ref{Fig:DRF_sf_p}(a), the out-of-phase (lower) gapless mode strongly responds to $\hat{F}_{1,i}$.
We can also see that the in-phase (upper) gapless mode and the high-momentum part of a gapful mode respond weakly to $\hat{F}_{1, i}$.
It has been pointed out that the low-momentum part of gapful modes does not respond to density fluctuations also in the case of the one-component Bose-Hubbard model~\cite{PRB_75_085106}.
Fig.~\ref{Fig:DRF_sf_p}(b) shows the dynamical structure factors for $\hat{F}^{\rm in}_{i}$.
For the fluctuation of this type, strongly responding modes are the in-phase (upper) gapless mode and the high-momentum part of the gapful mode that is the fifth excitation branch shown in Fig.~\ref{Fig:exc_sf_p}(c).
However, the out-of-phase (lower) gapless mode and the other gapful modes do not respond to the in-phase fluctuation.
Fig.~\ref{Fig:DRF_sf_p}(c) shows the dynamical structure factor for $\hat{F}^{\rm out}_{i}$.
In this case, the out-of-phase (lower) gapless mode responds strongly.
The fourth excitation mode also respond but this response is very weak.
\begin{figure}[hts]
\centering
\begin{tabular}{c}
\subfigure[]{\includegraphics[width=7.2cm,clip]{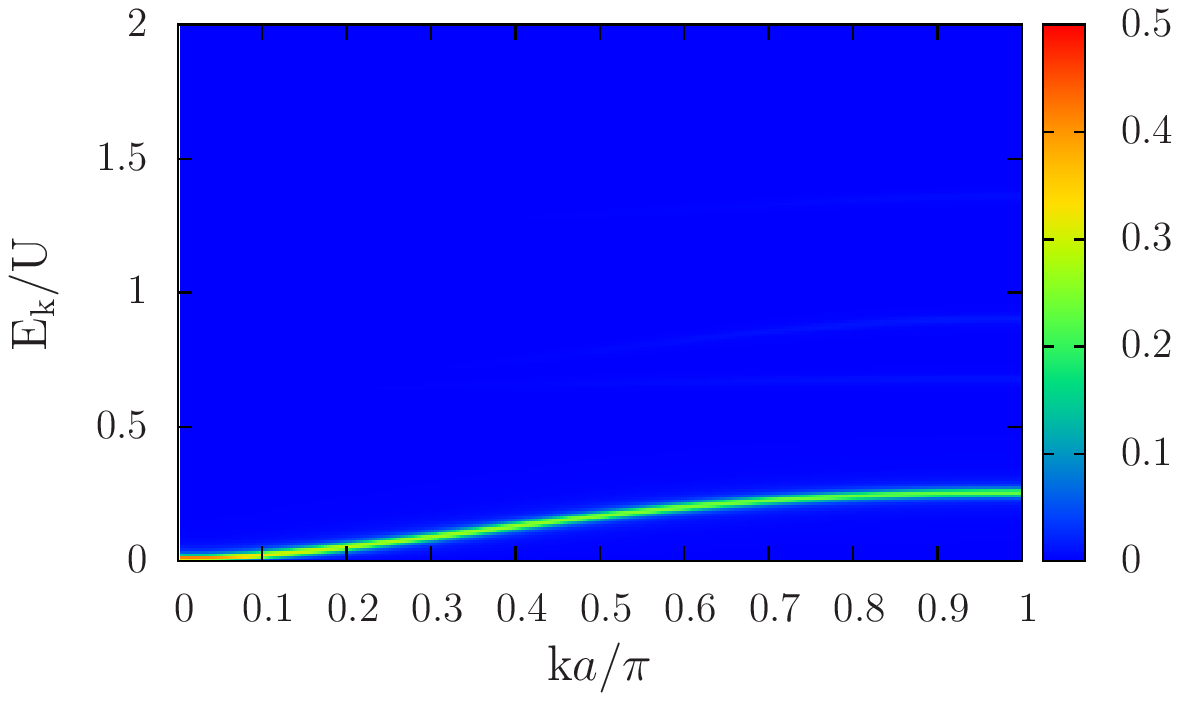}}\\
\end{tabular}
\begin{tabular}{cc}
\subfigure[]{\includegraphics[width=7.2cm,clip]{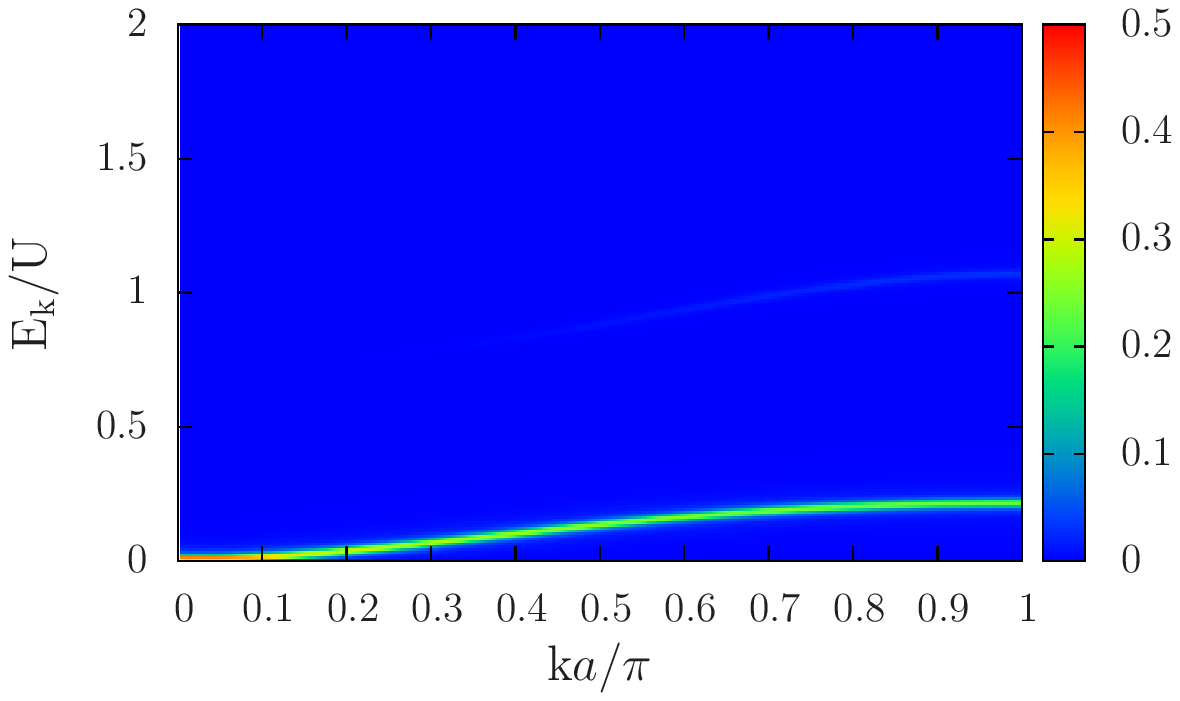}}
\subfigure[]{\includegraphics[width=7.2cm,clip]{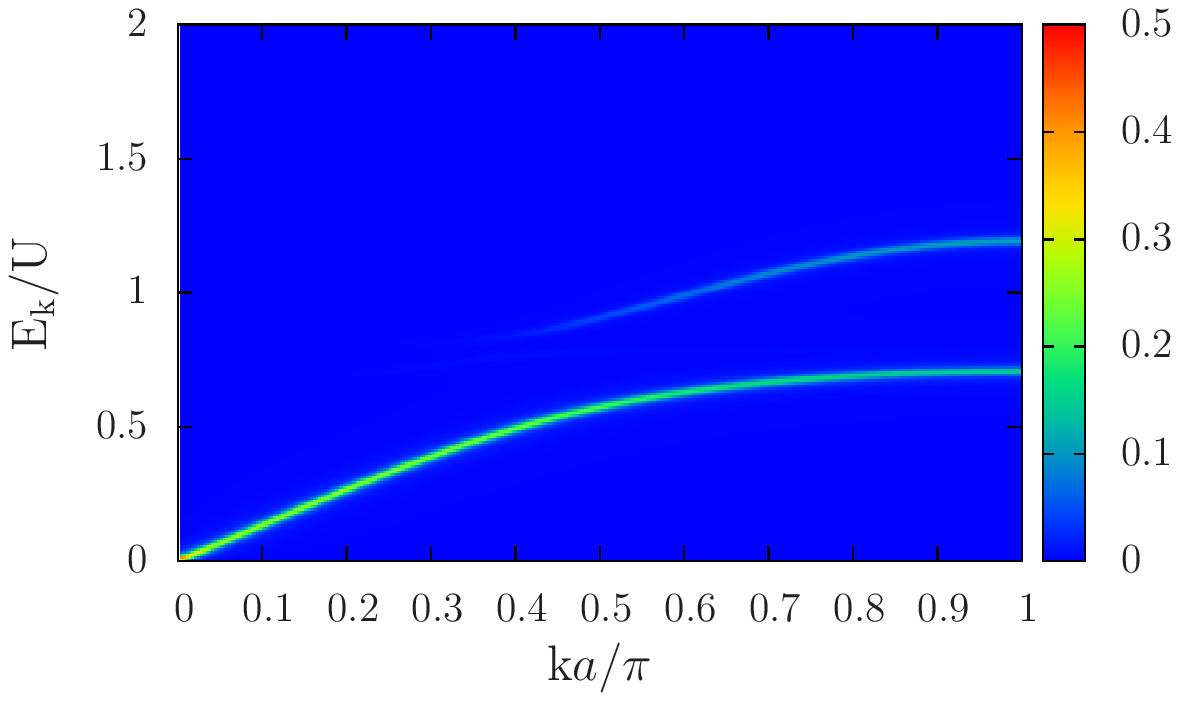}}
\end{tabular}
\caption{\label{Fig:DRF_sf_m}(Color online) Dynamical structure factors $\bar{S}_{\hat{F}}(\mathbf{k},\omega)$ of the SF phase with $n_{\rm t}=1$ for $t/U=0.07$ and $U_{12}/U=-0.7$. Shown are the responses to the density fluctuation of one-component $\hat{F}_{1,i}$ (a), the in-phase density fluctuation $\hat{F}^{\rm in}_i$ (b), and the out-of-phase density fluctuation $\hat{F}^{\rm out}_i$ (c).}
\end{figure}

In Fig.~\ref{Fig:DRF_sf_m}, we plot the dynamical structure factors for $n_{\rm t}=1$, $t/U=0.07$ and $U_{12}/U=-0.7$.
As shown in Fig.~\ref{Fig:DRF_sf_m}(a), the in-phase (lower) gapless mode strongly responds to the one-component fluctuation and no other mode significantly respond.
The modes responding to the in-phase or out-of-phase fluctuations are the in-phase (lower) or out-of-phase gapless mode and the high-momentum part of a gapful mode.

\begin{figure}[hts]%
\centering
\begin{tabular}{cc}
\subfigure[]{\includegraphics[clip,width=7.2cm,clip]{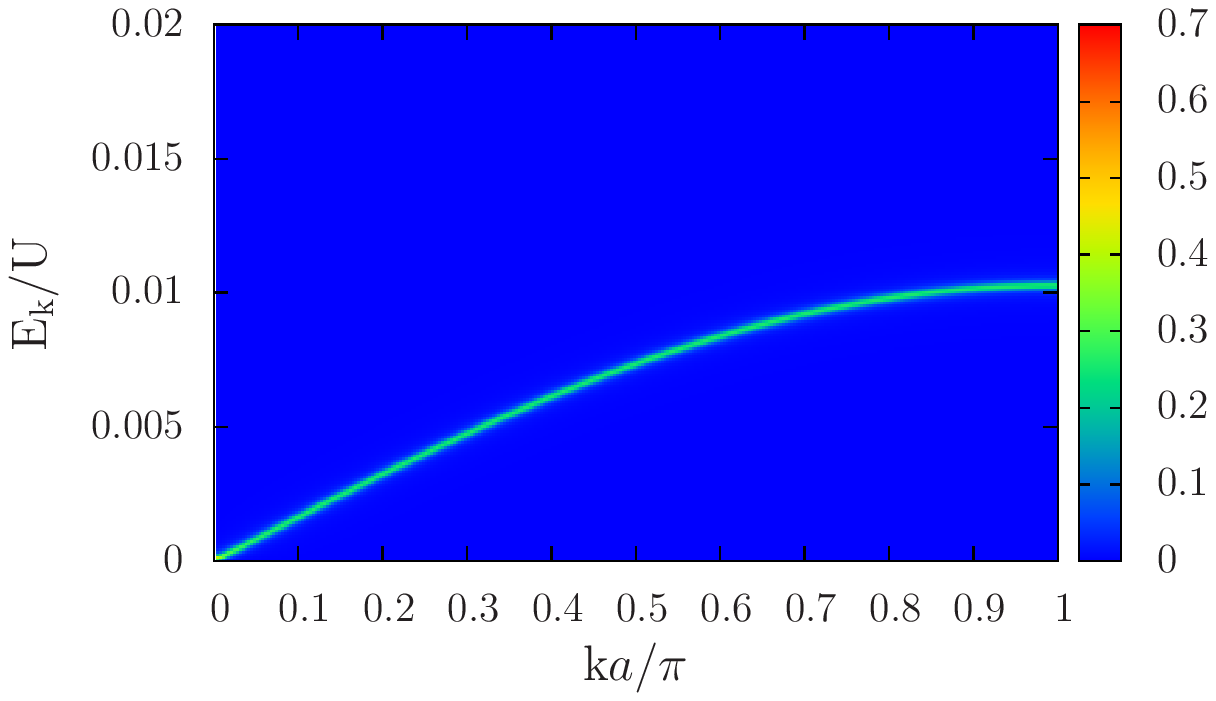}}
\subfigure[]{\includegraphics[clip,width=7.2cm,clip]{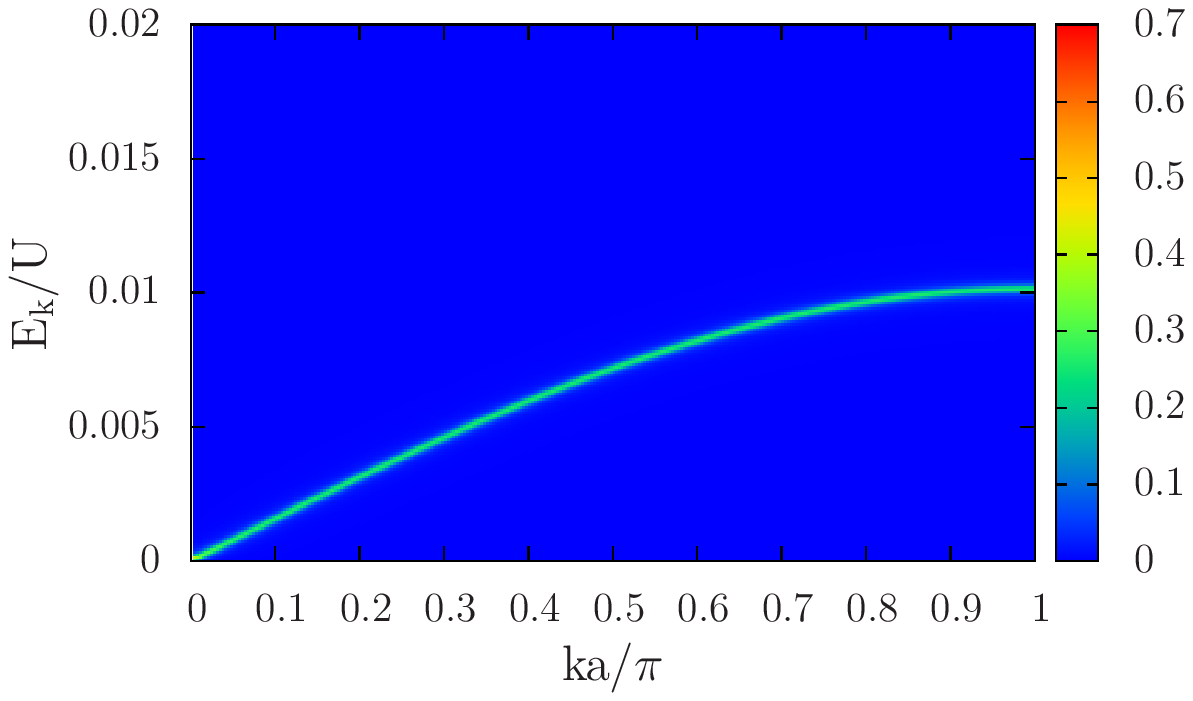}}
\end{tabular}
\caption{\label{Fig:DRF_eff}(Color online) Dynamical structure factors $\bar{S}_F(\mathbf{k},\omega)$ of (a) the PSF at $U_{12}/U=-0.7$ and $t/U=0.03$, and (b) the CFSF phase at $U_{12}/U=0.9$ and $t/U=0.03$.}
\end{figure}

We next calculate the dynamical structure factors for the PSF and CFSF phases.
In Fig.~\ref{Fig:DRF_eff}, we plot $S_{\hat{F}}({\bf k},\omega)$ with respect to the one-component density fluctuation $\hat{F}_{1, i}$.
We see that the gapless mode and no gapful mode respond to the one-component fluctuation for each phase.
In the PSF phase, gapless mode responds also to $\hat{F}^{\rm in}_i$, but does not respond to $\hat{F}^{\rm out}_i$.
On the other hand, in the CFSF phase, gapless mode responds also to $\hat{F}^{\rm out}_i$ and does not respond to $\hat{F}^{\rm in}_i$.
These results illuminate the essential property of the PSF (CFSF) state that the out-of-phase (in-phase) motion is forbidden while the in-phase (out-of-phase) motion exhibits superfluidity.

Since each phase exhibits different responses to the density fluctuations as shown above, measurement of the dynamical structure factors through the Bragg spectroscopy can be used to identify the phases.
A similar suggestion has been made in Ref.~\cite{PRA_84_041609}, where dynamic responses of the SF, PSF, and CFSF states to several types of trap displacement have been investigated.
We note that our results qualitatively explain the dynamics presented in Ref.~\cite{PRA_84_041609}.

\section{\label{sec6}Conclusions}
We have investigated ground state properties and excitations of Bose-Bose mixtures in an optical lattice using the Gutzwiller approximation.
We have obtained the ground-state phase diagrams in the $(\mu/U, zt/U)$ plane and found that the SF-to-MI phase transition can be of the first order in a wide range of $\mu/U$ for the strong inter-component interaction with even total fillings.
We have calculated the excitation spectra for the several phases by solving the linearized equations of motion.
In the SF phase, we have found a few gapful modes that are regarded as amplitude modes near the SF-to-MI transitions.
When $t/U$ decreases toward the second-order SF-to-MI transitions at $n_{\rm t}=1$ or $2$, one or two of the gapful modes descend and their low-energy part coincides with the gapless Bogoliubov modes at the transition point.
We have also calculated the excitation spectra in the PSF and CFSF phases, and shown that there is the gapless mode that supports superfluidity of pairs and anti-pairs, respectively.
Moreover, we have computed the dynamical structure factors by means of the linear response theory.
We have studied the responses to three types of density fluctuation for each quantum phase.
We have shown that branches which respond to the density fluctuations are different in each quantum phase. These results suggest that the quantum phases should be identified by measuring the responses to the density fluctuations.

One of the most characteristic properties of first-order transitions is the hysteresis, where the phase transition point depends upon its history, namely from which phase we start changing parameters to cause the transition.
Recent theoretical work has studied quantum phase transitions in the hardcore Bose-Hubbard model with long-range interactions on a triangular lattice and predicted that when the chemical potential is varied, the hysteresis associated with the first order SF-to-MI transition exhibits an anomalous behavior in which a standard hysteresis loop is not formed~\cite{PRA_85_021601}.
It will be interesting to study the hysteresis in the two-component Bose-Hubbard model in order to reveal whether or not such an anomalous behavior is present also in the absence of long-range interactions.

\begin{acknowledgments}
We gratefully acknowledge valuable comments and discussions with E. Arahata, M. Kanetaka, L. Mathey, T. Saito, D. Schneble, S. Tsuchiya, and D. Yamamoto.
\end{acknowledgments}

\appendix
\section{Linear response theory within the Gutzwiller approximation}
\label{sec:appA}
For Simplicity, we use the contraction basis $|n\rangle\equiv|n_1,n_2\rangle$ and wavefunction $|\Psi\rangle\equiv\prod_i\sum_nf_n^{(i)}|n\rangle_i$ in this appendix.
The equation of motion for the coefficient $f_n^{(i)}(\tau)$ can be derived by imposing the stationary conditions on the effective action $S=\int d\tau \left\langle \Psi\left|{\rm{i}}\hbar\frac{d}{d\tau}-\hat{H}-\hat{H}_{pert}(\tau)\right|\Psi\right\rangle$ with respect to $f_n^{(i)\ast}$:
\begin{align}
\label{eqn:a7}
{\rm{i}}\hbar\frac{d}{d\tau}f_n^{(i)}=\frac{\partial E}{\partial f_n^{(i)\ast}}-\sum_{m}f_m^{(i)}\left( \lambda_iG_{n,m} e^{-{\rm{i}}\omega\tau}e^{\eta\tau}+\lambda_i^\ast G_{m,n}^\ast e^{{\rm{i}}\omega\tau}e^{\eta\tau} \right).
\end{align}
Here, we define $E\equiv\langle\Psi|\hat{H}|\Psi\rangle$.
$\hat{H}_{pert}(\tau)$ and $G_{m,n}$ have been defined as (\ref{eqn:a1}) and (\ref{eqn:a2}), respectively.
In order to consider the small amplitude oscillations, we assume the following form:
\begin{align}
\label{eqn:a8}
f_n^{(i)}(\tau)=\left[\tilde{f}_n^{(i)}+\delta f_n^{(i)}(\tau)\right]e^{-{\rm{i}}\tilde{\omega}_i \tau},
\end{align}
where $\tilde{f}_n^{(i)}$ is the stationary solution in the absence of the external perturbation, which satisfies
\begin{align}
\label{eqn:a9}
\hbar\tilde{\omega}_i\tilde{f}_n^{(i)}=\left.\frac{\partial E}{\partial f_n^{(i)\ast}}\right|_{f_n^{(i)}=\tilde{f}_n^{(i)}}.
\end{align}
Thus, the linearized equation of motion for $\delta f_n^{(i)}$ is given by
\begin{align}
\label{eqn:a3}
{\rm{i}}\hbar\frac{d\delta f_n^{(i)}}{d\tau}=&-\hbar\tilde{\omega}_i\delta f_n^{(i)}(\tau)+\sum_j\sum_m\left(\left.\frac{\partial^2E}{\partial f_m^{(j)}\partial f_n^{(i)\ast}}\right|_{f_n^{(i)}=\tilde{f}_n^{(i)}} \delta f_m^{(j)} + \left.\frac{\partial^2E}{\partial f_m^{(j)\ast}\partial f_n^{(i)\ast}}\right|_{f_n^{(i)}=\tilde{f}_n^{(i)\ast}}\delta f_m^{(j)\ast}\right) \notag\\
&-\sum_{m}\tilde{f}_m^{(i)}\left( \lambda_iG_{n,m} e^{-{\rm{i}}\omega\tau}e^{\eta\tau}+\lambda_i^\ast G_{m,n}^\ast e^{{\rm{i}}\omega\tau}e^{\eta\tau} \right).
\end{align}
We assume the fluctuation in the following form:
\begin{align}
\label{eqn:a14}
\delta f_n^{(i)}(\tau)=u_{i,n}e^{-{\rm{i}}\omega\tau}e^{\eta\tau}-v_{i,n}^\ast e^{{\rm{i}}\omega\tau}e^{\eta\tau},
\end{align}
where the Bogoliubov amplitude $u_{i,n}$ and $v_{i,n}$ are time-independent.
Inserting Eq.~(\ref{eqn:a14}) to (\ref{eqn:a3}), we obtain
\begin{align}
\label{eqn:a17}
\left( \hbar\omega+{\rm{i}}\hbar\eta +\hbar\tilde{\omega}_i \right)u_{i,n}=&\sum_j\sum_m\left( \left.\frac{\partial^2E}{\partial f_m^{(j)}\partial f_n^{(i)\ast}}\right|_{f_n^{(i)}=\tilde{f}_n^{(i)}} u_{j,m}-\left.\frac{\partial^2E}{\partial f_m^{(j)\ast}\partial f_n^{(i)\ast}}\right|_{f_n^{(i)}=\tilde{f}_n^{(i)\ast}}v_{j,m} \right)\notag\\
&-\sum_m\tilde{f}_m^{(i)}\lambda_iG_{n,m},
\end{align}
and
\begin{align}
\label{eqn:a19}
-\left(\hbar\omega+{\rm{i}}\hbar\eta -\hbar\tilde{\omega}_i\right)v_{i,n}=&\sum_j\sum_m\left( \left.\frac{\partial^2E}{\partial f_m^{(j)\ast}\partial f_n^{(i)}}\right|_{f_n^{(i)}=\tilde{f}_n^{(i)\ast}}v_{j,m}-\left.\frac{\partial^2E}{\partial f_m^{(j)}\partial f_n^{(i)}}\right|_{f_n^{(i)}=\tilde{f}_n^{(i)}} u_{j,m}  \right)\notag\\
&+\sum_m\tilde{f}_m^{(i)\ast}\lambda_i G_{m,n}.
\end{align}

We next expand the fluctuation $\delta f_n^{(i)}$ in terms of the normal-modes as
\begin{align}
\label{eqn:a20}
u_{i,n}=&\sum_\nu\left( A_\nu u_{i,n}^{(\nu)}-B_\nu^\ast v_{i,n}^{(\nu)\ast} \right),\\
\label{eqn:a21}
v_{i,n}=&\sum_\nu\left( A_\nu v_{i,n}^{(\nu)}-B_\nu^\ast u_{i,n}^{(\nu)\ast} \right),
\end{align}
where $u_{i,n}^{(\nu)},v_{i,n}^{(\nu)}$ are the solutions of the Bogoliubov Eq.~(\ref{eqn:GW-bogo_1}), and $\nu$ is the mode index.
We assume the Bogoliubov amplitude $u_{i,n}^{(\nu)}$ and $v_{i,n}^{(\nu)}$ are normalized by $\sum_i\sum_n\left( |u_{i,n}^{(\nu)}|^2-|v_{i,n}^{(\nu)}|^2 \right)=1$.
With this normalization, they satisfy the orthogonality relation $\sum_i\sum_n\left( u_{i,n}^{(\nu)\ast}u_{i,n}^{(\nu^\prime)}-v_{i,n}^{(\nu)\ast}v_{i,n}^{(\nu^\prime)} \right)=\delta_{\nu,\nu^\prime}$.
From Eqs.~(\ref{eqn:a17}), (\ref{eqn:a19}), (\ref{eqn:a20}) and (\ref{eqn:a21}), we obtain
\begin{align}
\label{eqn:a26-1}
\sum_\nu\left[ \hbar\left( \omega+{\rm{i}}\eta-\omega_\nu \right)u_{i,n}^{(\nu)}A_\nu-\hbar\left( \omega+{\rm{i}}\eta+\omega_\nu \right)v_{i,n}^{(\nu)\ast}B_\nu^\ast\right]=&-\sum_m\tilde{f}_m^{(i)}\lambda_iG_{n,m},\\
\label{eqn:a26-2}
\sum_\nu\left[ \hbar\left( \omega+{\rm{i}}\eta-\omega_\nu \right)v_{i,n}^{(\nu)}A_\nu-\hbar\left( \omega+{\rm{i}}\eta+\omega_\nu \right)u_{i,n}^{(\nu)\ast}B_\nu^\ast \right]=&-\sum_m\tilde{f}_m^{(i)\ast}\lambda_i G_{m,n}.
\end{align}
From Eqs.~(\ref{eqn:a26-1}) and (\ref{eqn:a26-2}), we obtain the solutions for $A_\nu$ and $B_\nu$ as
\begin{align}
\label{eqn:a35}
A_\nu=&-\frac{1}{\hbar}\sum_i\frac{\langle \nu|G|0\rangle_i}{\omega+{\rm{i}}\eta-\omega_\nu}\lambda_i=-\frac{1}{\hbar}\sum_i\frac{\langle 0|G^\dag|\nu\rangle_i^\ast}{\omega+{\rm{i}}\eta-\omega_\nu}\lambda_i,\\
\label{eqn:a36}
B_\nu^\ast=&\frac{1}{\hbar}\sum_i\frac{\langle 0|G|\nu\rangle_i}{\omega+{\rm{i}}\eta+\omega_\nu}\lambda_i=\frac{1}{\hbar}\sum_i\frac{\langle\nu|G^\dag|0\rangle_i^\ast}{\omega+{\rm{i}}\eta+\omega_\nu}\lambda_i,
\end{align}
where we have introduced the compact notations of the matrix element as
\begin{align}
\label{eqn:a33}
\langle \nu|G|0\rangle_i\equiv&\sum_n\sum_m\left(u_{i,n}^{(\nu)\ast}G_{n,m}\tilde{f}_m^{(i)}-\tilde{f}_m^{(i)\ast} G_{m,n}v_{i,n}^{(\nu)\ast} \right),\\
\langle 0|G|\nu\rangle_i\equiv&\sum_n\sum_m\left(\tilde{f}_m^{(i)\ast} G_{m,n}u_{i,n}^{(\nu)}-v_{i,n}^{(\nu)}G_{n,m}\tilde{f}_m^{(i)} \right),\\
\label{eqn:a34}
\langle \nu|G^\dag|0\rangle_i\equiv&\sum_n\sum_m\left(u_{i,n}^{(\nu)\ast}G_{n,m}^\ast\tilde{f}_m^{(i)\ast}-\tilde{f}_m^{(i)} G_{m,n}^\ast v_{i,n}^{(\nu)\ast} \right)=\langle 0|G|\nu\rangle_i^\ast,\\
\langle 0|G^\dag|\nu\rangle_i\equiv&\sum_n\sum_m\left(\tilde{f}_m^{(i)} G_{m,n}^\ast u_{i,n}^{(\nu)\ast}-v_{i,n}^{(\nu)\ast}G_{n,m}^\ast\tilde{f}_m^{(i)^\ast} \right)=\langle \nu|G|0\rangle_i^\ast.
\end{align}

Having obtained the linearized solution for $\delta f_n^{(i)}$, we now calculate the fluctuation of the physical quantity associated with $\hat{F}_i$:
\begin{align}
\label{eqn:a37}
\langle \hat{F}_i \rangle=\sum_{n,m}f_n^{(i)\ast}F_{n,m}f_n^{(i)}\approx\langle \hat{F}_i \rangle_0+\delta\langle \hat{F}_i \rangle,
\end{align}
where
\begin{align}
\label{eqn:a38}
\langle \hat{F}_i \rangle_0\equiv&\sum_{n,m}\tilde{f}_n^{(i)\ast}F_{n,m}\tilde{f}_n^{(i)},\\
\delta\langle \hat{F}_i \rangle\equiv&\sum_{n,m}\left( \tilde{f}_n^{(i)\ast}\delta f_m^{(i)} +\delta f_n^{(i)\ast}\tilde{f}_m^{(i)}\right)F_{n,m}.
\end{align}
Using Eqs.~(\ref{eqn:a14}), (\ref{eqn:a20}), (\ref{eqn:a21}), (\ref{eqn:a35}), and (\ref{eqn:a36}), we obtain
\begin{align}
\label{eqn:a41}
\delta\langle \hat{F}_i \rangle=&-\frac{1}{\hbar}\sum_\nu\sum_j\left[\left( \frac{\langle 0|F|\nu\rangle_i\langle \nu|G|0\rangle_j}{\omega+{\rm{i}}\eta-\omega_\nu}-\frac{\langle 0|G|\nu\rangle_j\langle \nu|F|0\rangle_i}{\omega+{\rm{i}}\eta+\omega_\nu} \right)\lambda_je^{-{\rm{i}}\omega\tau}e^{\eta\tau}\right.\notag\\
&\left.+\left(\frac{\langle 0|G^\dag|\nu\rangle_j\langle \nu|F|0\rangle_i}{\omega-{\rm{i}}\eta-\omega_\nu}-\frac{\langle 0|F|\nu\rangle_i\langle \nu|G^\dag|0\rangle_j}{\omega-{\rm{i}}\eta+\omega_\nu} \right)\lambda_je^{{\rm{i}}\omega\tau}e^{\eta\tau} \right]
\end{align}
This can be written in terms of the linear response function as
\begin{align}
\label{eqn:a42}
\delta\langle \hat{F}_i \rangle=\sum_j\left[ \chi_{F,G}(i,j,\omega)\lambda_je^{-{\rm{i}}\omega\tau}e^{\eta\tau}+ \chi_{F,G^\dag}(i,j,-\omega)\lambda_je^{{\rm{i}}\omega\tau}e^{\eta\tau} \right],
\end{align}
where
\begin{align}
\label{eqn:a43}
\chi_{F,G}(i,j,\omega)\equiv-\frac{1}{\hbar}\sum_\nu\left( \frac{\langle 0|F|\nu\rangle_i\langle \nu|G|0\rangle_j}{\omega+{\rm{i}}\eta-\omega_\nu}-\frac{\langle 0|G|\nu\rangle_j\langle \nu|F|0\rangle_i}{\omega+{\rm{i}}\eta+\omega_\nu} \right).
\end{align}
Using the Fourier transformation, Eqs.~(\ref{eqn:a42}) and (\ref{eqn:a43}) are written as
\begin{align}
\delta\langle \hat{F}_i \rangle=\sum_{\mathbf{k}}\left[ \chi_{F,G}(\mathbf{k},\omega)e^{-{\rm{i}}\omega t}e^{\eta t}+ \chi_{F,G^\dag}(\mathbf{k},-\omega)e^{{\rm{i}}\omega t}e^{\eta t} \right]e^{{\rm{i}}\mathbf{k}\cdot\mathbf{r}_i}\lambda_{\mathbf{k},\omega},
\end{align}
and
\begin{align}
\chi_{\hat{F},\hat{G}}(\mathbf{k},\omega)=-\frac{1}{\hbar}\sum_\lambda\left[ \frac{\langle 0|\hat{F}|\lambda\rangle\langle\lambda|\hat{G}|0\rangle}{\omega+{\rm{i}}\eta-\omega_{\mathbf{k},\lambda}} - \frac{\langle 0|\hat{G}|\lambda\rangle\langle\lambda|\hat{F}|0\rangle}{\omega+{\rm{i}}\eta+\omega_{\mathbf{k},\lambda}} \right].
\end{align}

\end{document}